\documentclass{aa}  
\usepackage{graphicx}
\usepackage{txfonts}
\usepackage{multirow}
\usepackage{xcolor,colortbl}
\usepackage{subcaption}
\newcommand{\msun}{$\mathrm{M}_{\odot}$}
\newcommand{\hinv}{$h^{-1}$}

\newcommand{\tre}{{\tt Three Hundred}}

\begin{document}

   \title{The Three Hundred Project: Modeling Baryon and Hot-Gas Fraction Evolution in Simulated Clusters}

   \author{Elena Rasia\fnmsep\thanks{\email{elena.rasia@inaf.it}}
    \inst{1,2,3}
    \and
    Roberta Tripodi
   \inst{1,2,4}
    \and
    Stefano Borgani\inst{1,2,5,6,7}
    \and
    Veronica Biffi\inst{1,2}
    \and
    Camille Avestruz\inst{3}
    \and
    Weiguang Cui\inst{8,9}
    \and 
    Marco De Petris\inst{10}
    \and
    Klaus Dolag\inst{11,12}
    \and
    Dominique Eckert\inst{13}
    \and
    Stefano Ettori\inst{14,15}
    \and
    Massimo Gaspari\inst{16}}

   \institute{
INAF -- Osservatorio Astronomico di Trieste, via Tiepolo 11, I-34131, Trieste, Italy 
\and
IFPU, Institute for Fundamental Physics of the Universe, Via Beirut 2, 34014 Trieste, Italy
\and 
Department of Physics; University of Michigan, Ann Arbor, MI 48109, USA
\and
University of Ljubljana, Faculty of Mathematics and Physics, Jadranska ulica 19, SI-1000 Ljubljana, Slovenia
\and
Dipartimento di Fisica dell'Universit\`a di Trieste, Sez. di Astronomia, via Tiepolo 11, I-34131 Trieste, Italy
\and
Center on High Performance Computing, Big Data and Quantum Computing, via Magnanelli 2, 40033, Casalecchio di Reno, Italy
\and
INFN, Instituto Nazionale di Fisica Nucleare, Via Valerio 2, I-34127, Trieste, Italy
\and
Departamento de Física Te\'orica, M-8, Universidad Aut\'onoma de Madrid, Cantoblanco E-28049, Madrid, Spain
\and
Centro de Investigaci\'on Avanzada en F\'isica Fundamental (CIAFF), Universidad Aut\'noma de Madrid, Cantoblanco, E-28049 Madrid, Spain
\and
Dipartimento di Fisica, Sapienza Università di Roma, Piazzale Aldo Moro 5, I-00185 Roma, Italy;
\and
Universit\"ats-Sternwarte, Fakult\"at f\"ur Physik, Ludwig-Maximilians-Universit\"at M\"unchen, Scheinerstr.1, 81679
\and
Max-Plank-Institut f\"ur Astrophysik, Karl-Schwarzschild Strasse 1, D-85740 Garching, Germany
\and
Department of Astronomy, University of Geneva, ch. d'Ecogia 16, CH-1290 Versoix, Switzerland
\and
INAF, Osservatorio di Astrofisica e Scienza dello Spazio, via Piero Gobetti 93/3, 40129 Bologna, Italy 
\and
INFN, Sezione di Bologna, viale Berti Pichat 6/2, 40127 Bologna, Italy 
\and
Department of Physics, Informatics and Mathematics, University of Modena and Reggio Emilia, 41125 Modena, Italy
}

 %  \date{Received September 15, 1996; accepted March 16, 1997}

% \abstract{}{}{}{}{} 
% 5 {} token are mandatory
 
  \abstract
   {The baryon fraction of galaxy clusters, expressed as the ratio between the mass in baryons (including both stars and cold/hot gas) and the total mass is a powerful tool to inform on the cosmological parameters while the hot-gas fraction provides indications on the physics of the intracluster plasma and its interplay with the processes driving galaxy formation.  
   }
   {Using cosmological hydrodynamical simulations of about 300 simulated massive galaxy clusters with median mass $M_{500}\approx7 \times 10^{14}$M$_{\odot}$ at $z=0$, we model the relations between total mass and either baryon fraction or the hot gas fractions at overdensities $\Delta = 2500$, $500$, and $200$ with respect to the cosmic critical density, and their evolution from $z\sim 0$ to $z\sim 1.3$. 
   }
   {We utilize the simulated galaxy clusters from the \tre\ project which include star formation and feedback from both supernovae and active galactic nuclei. We fit the simulation results for such scaling relations against three analytic forms (linear, quadratic, and logarithmic in a logarithmic plane) and three forms for the redshift dependence, considering as a variable both the inverse of cosmic scale factor, $(1+z)$, and the Hubble expansion rate, $E(z)$.
   }
   {We show that power-law dependencies on cluster mass poorly describe the investigated relations.  A power-law fails to simultaneously capture the flattening of the total baryon and gas fractions at high masses, their drop at the low masses, and the transition between these two regimes. The other two functional forms provide a more accurate description of the curvature in mass scaling. The fractions measured within smaller radii exhibit a stronger evolution than those measured within larger radii.
   }
  {From the analysis of these simulations, we evince that as long as we include systems in the mass range herein investigated, the baryon or gas fraction can be accurately related to the total mass through either a parabola or  a logarithm in the logarithmic plane. The trends are common to all modern hydro simulations, although the amplitude of the drop at low masses might differ. Being able to observationally determine the gas fraction in groups will, thus, provide constraints on the baryonic physics.
  }
   \maketitle
%\ref
%sec:intro    sec:sim  sec:met
%sec:res      sec:concl

%--------------------------------------------
%--------------------------------------------
\section{Introduction}
\label{sec:intro}

Precise measurements of the baryon fraction within galaxy clusters and groups have the twofold scope of informing about cosmological parameters \citep{allen.etal.2011,mantz.etal.2014, Mantz.etal.2022} and about the non-gravitational processes such as heating and cooling that govern the baryon physics during clusters' formation and evolution \citep[see reviews by][]{eckert.etal.2021,Oppenheimer.etal.21,lovisari_maughan22}. 

Cosmological studies are based on samples of massive clusters, whose baryon-to-total mass ratio is very close to the cosmic baryon fraction: $f_{b,cosmic} = \Omega_b/\Omega_M$. 
Therefore, the matter cosmic density parameter, $\Omega_M$, can be inferred from the observational measurements of the cluster baryon fraction once a certain value for the Hubble parameter is assumed \citep[][and references therein]{Mantz.etal.2022} and the cosmic baryon density, $\Omega_b$, is considered as from cosmic-microwave-background studies \citep[e.g.,][]{Planck2018} or Big Bang nucleosynthesis models \citep[e.g.][]{fields.etal.2020}. In addition, measuring the baryon fraction as a function of redshift informs on the expansion history of the universe and thus it is also sensitive to the dark-energy density parameter, $\Omega_{\Lambda}$ and, possibly, to its equation of state \citep{ettori.etal.2009}.

Studies of numerical simulations also promote this approach based on the fact that the {\it overall} baryon fraction predicted in {\it massive} clusters is weakly sensitive to the details of the physical processes included in the simulations - e.g. related to the models of star formation, and of both stellar and active-galactic-nuclei (AGN) feedback - which in general determine the observational properties of the intracluster medium \citep[ICM,][]{planelles.etal.2013,battaglia.etal.2013,Gaspari2020,cui.etal.2022}. On the other hand, such astrophysical processes determine the exact amount of stellar and gas fractions which, in massive systems are tightly anti-correlated: systems with lower gas fraction tend to have higher stellar fraction and vice versa \citep{wu.etal.2015,truong.etal.2018,li.etal.2023}. This theoretical prediction has been confirmed using observations of a sample of 41 X-ray bright clusters, concluding that massive clusters can be considered as `closed boxes' within a fair approximation \citep{farahi.etal.2018}.

While the simplicity of these arguments supports the use of the baryon fraction as a cosmological probe, they are unfortunately valid only for massive clusters \citep[e.g.][]{akino.etal.2022}. As one shifts towards galaxy groups, the gas fraction decreases by an amount which is larger than the corresponding increase of the stellar fraction, thus leading to a net decrease of the total baryon fraction \citep{giodini.etal.2009}. This observational fact, linked to the break of the scaling relations between the total mass and various gas-related properties, such as X-ray luminosity or SZ signal, has been known and discussed since some time in the literature \citep[see review by ][and references therein]{lovisari.etal.2021}. The main culprit of the deviation of the baryon fraction (and particularly the gas fraction) from the cosmic expectation at the galaxy-group scale, is the cumulative activity of the central AGN. Indeed, during the early phases of the cluster assembly at $z\gtrsim 2$ the AGN feedback can easily expel the central gas from the then shallower potential well of the cluster progenitors, hampering at the same time the star-formation process \citep{fabjan.etal.2010,mccarthy.etal.2007, mccarthy.etal.2010,Gaspari2014,biffi.etal.2018,henden.etal.2018}. For this reason, systems in the local universe with masses $M_{500}\sim 10^{14}$M$_{\odot}$ present some level of baryon depletion \citep{ettori.etal.2006, angelinelli.etal.2023, ayromlou.etal.2023}, which can significantly increase in the group regime,  $M_{500}\simeq (10^{13}-10^{14})$M$_{\odot}$ \citep{davies.etal.2020,Oppenheimer.etal.21,donahue_voit, popesso.etal.2024arxiv}. 

Connected to this aspect, while on the one hand the relations between the gas or stellar mass and the total mass typically were thought to have a power-law shape of the form $M \propto M_{\rm gas}^{\alpha}$ or $M\propto M_{\rm star}^{\beta}$, because of the expected self-similar behavior of the scaling relations of massive clusters \citep{kaiser86}, on the other hand, recent works exploring masses from $10^{13}M_{\odot}$ to above $10^{15}M_{\odot}$ challenged this simplistic description and have proposed that the slope of the scaling relation changes from groups to clusters  \citep{lebrun.etal.2017, truong.etal.2018,farahi.etal.2018,Gaspari2019,pop.etal.2022}.

In this work, we will limit our study to the baryon and hot-gas fractions for a large sample of $z\sim 0$ simulated {\it massive galaxy clusters} and focus on their evolution showing that precautions need to be taken at higher redshifts when clusters are smaller. 
Specifically,  we propose and test simple analytical fitting functions that go beyond the power-law description and could, in the future, be easily extended to the group scale.
This paper differs from previous investigations on the baryon and hot gas fraction in two main aspects: 
$(i)$ we compare various analytic expressions to fit the simulated data, similarly to \cite{lebrun.etal.2014} and \cite{pop.etal.2022} (see also \citealt{miller.etal.2025}) but using simpler descriptions and considering a total of about 20 different analytic expressions which capture the dependence on mass, overdensity and redshift;
$(ii)$ the set of simulated clusters analyzed here includes hundreds of massive halos, with 95 percent of them having virial masses\footnote{For the cosmology considered in the simulations (see Sect.~\ref{sec:sim}), the virial radius defined as in \cite{bryan_norman} is close to the overdensity of 100, which is the exact value used here.} above $10^{15}M_\odot$ at $z=0$ thus providing us with a statistically robust high-mass anchor at $z=0$ which we are able to follow up to $z=1.32$.
Indeed, the simulation suite on which our analysis is based, the \tre\ is one of the largest samples of massive systems simulated with AGN feedback and reaching good numerical resolution. Its large statistics of massive clusters is comparable to {\tt FLAMINGO-L1} \citep{schaye.etal.2023}, {\tt TNG-Clusters} \citep{nelson.etal.2023} and {\tt MACSIS} \citep{barnes.etal.macsis.2017}, and only surpassed, albeit at lower resolution, by  {\tt MAGNETICUM-Box0}\footnote{http://www.magneticum.org/simulations.html} and {\tt FLAMINGO-L2}. At the same time, other simulations with high particle-mass resolution have only a sparse number of objects with $M_{500}>10^{15}$\msun, e.g,. {\tt DIANOGA} (\citealt{rasia.etal.2015,bassini.etal.2020}, Esposito et al. 2025, in preparation), {\tt C-EAGLES} \citep{barnes.etal.2017ceagle}, {\tt Hydrangea} \citep{bahe.etal.2017}, {\tt Illustris-TNG} \citep{pillepich.etal.2018}, {\tt MilleniumTNG} \citep{pakmor.etal.2023}. Thanks to the large statistics of clusters that we can follow from $z=0$ to high-redshift, we have more than 250  well-resolved halos even at $z=1.32$ with a mass range from $M_{500}=4 \times 10^{13}$M$_{\odot}$ to $4\times 10^{14}$M$_{\odot}$.

The paper is organized as follows. In Sect.~\ref{sec:sim}, we introduce the simulations and the 
cluster redshift subsamples and the combined cluster sample. The best-fit functions and the fitting procedure are explained in Sect.~\ref{sec:met}. We compare the simulated data with observational samples in Sect.~\ref{sec:obse}. The baryon-fraction results derived from the redshift subsamples and for the combined sample, including all simulated objects, are presented in Sects.~\ref{sec:res_m} and \ref{sec:res_mz}, respectively. The analysis of the hot-gas baryon fraction is reported in Sect.~\ref{sec:gas}. We compare with other simulation results in Sect.~\ref{sec:compasim} and discuss in Sect.~\ref{sec:discussion} the differences with the other version of the \tre, carried out with {\tt GIZMO-SIMBA}, and with {\tt MAGNETICUM}, which shares several features with our code. The main conclusions are summarized in Sect.~\ref{sec:concl}. 
In Appendix~\ref{app:bar} we discuss of the  best-fitting results in relation to the baryon fraction and provide all numerical values of the parameters  (Tables~\ref{tab:eq1}-\ref{tab:eq3}); in Appendix~\ref{app:gas} we list the best-fit parameters for the hot gas fraction (Tables~\ref{tab:eq1gas}-\ref{tab:eq3gas}); in Appendix~\ref{app:evpar} we compare the two expressions used to model the redshift evolution; in Appendix~\ref{app:c} we present the results from some alternative functional forms and from independent redshift subsamples; finally, in Appendix~\ref{app:med} we discuss the evolution of the medians of both baryon and hot gas fractions for each redshift subsample.

Throughout the paper, the logarithms are always intended as decimal. 

%--------------------------------------------
%--------------------------------------------
   
%--------------------------------------------
%--------------------------------------------
\section{Simulations}
\label{sec:sim}
%--------------------------------------------
%--------------------------------------------
The \tre\ project \citep[hereafter The300,][]{cui.etal.2018} consists of a set of simulated regions centered on massive galaxy clusters. The sample is built following the `zoom-in' technique that extracted 324 Lagrangian regions around the most massive clusters identified within the MultiDark-Planck2 box\footnote{The MultiDark simulations are publicly available at the https://www.cosmosim.org database.} \citep{klypin.etal.2016} and resimulated them by adding baryonic physics. The parent simulation evolves only the dark-matter (DM) component with $3840^3$ DM particles in a cosmological box of $1.5$ comoving Gpc as a side. This simulation assumes the cosmological parameters derived by the Planck Collaboration \citep{Planck2016}: $h=0.6777$ for the reduced Hubble parameter, $n=0.96$ for the primordial spectral index, $\sigma_8=0.8228$ for the amplitude of the mass density fluctuations in a sphere of 8 \hinv Mpc comoving radius, and $\Omega_{\Lambda}=0.692885$, $\Omega_{\rm m}=0.307115$, and $\Omega_{\rm b}=0.048206$, respectively for the cosmological density parameters already mentioned in the Introduction. The re-simulated regions are evolved under the same cosmological model and therefore they are characterized by a cosmic baryon fraction of
\begin{equation}
\label{eq:fbcosmo}
f_{\mathrm b,\mathrm{cosmic}}=\frac{\Omega_{\rm b}}{\Omega_{\rm m}}\simeq 0.157. 
\end{equation}

The highest resolution region covers a radius of about $15$ \hinv comoving Mpc centered on the $z=0$ target halo, while the influence of the external tidal fields is traced by low-resolution particles. The high-resolution dark-matter particle mass is equal to $m_{\rm DM}=1.27 \times 10^9$ \hinv \msun, while the initial gas-particle mass is  $m_{\rm gas}=2.36 \times 10^8$ \hinv \msun.

This work is based on the resimulations carried out with an updated version of the {\tt GADGET-2} \citep{Springel2005} code, referred by The300 collaboration as {\tt GADGET-X}. This is based on the implementation of SPH as in \cite{beck.etal.2016}, and implementations of cooling, star formation, and feedback from stars and AGN similar to \cite{rasia.etal.2015}. Specifically, the uniform time-dependent UV background and the radiative cooling dependence on metallicity are as in \citet{Wiersma2009}; the star formation and thermal feedback from supernovae closely follow the original implementation by \citet{Springelhernquist2003}  and are connected to a detailed stellar evolution and chemical enrichment model as in \citet[][see also, \citealt{biffi.etal.2017,biffi.etal.2018} and \citealt{truong.etal.2019}]{Tornatore2007}. Finally, the hot and cold gas accretion onto super-massive black holes (SMBHs) powers the AGN feedback following the model by \cite{Steinborn2015}. 
Black holes are seeded with mass $M_{\rm BH}= 5 \times 10^6 h^{-1}$ M$_{\odot}$ in the center of halos when their mass is above $2.5 \times 10^{11} h^{-1}$ M$_{\odot}$. SMBHs grow through mergers or gas accretion \citep{bassini.etal.2019}. The latter is capped by the minimum of the Eddington limit and the $\alpha$-modified Bondi accretion rate, where $\alpha$ is a parameter fixed at 10 or 100 for the hot ($T>5 \times 10^5 K$) or the cold gas ($T<5\times 10^5K$) respectively \citep{gaspari.etal.2015}. The efficiency of the radiative and mechanical feedback  depends on both the gas accretion rate and the SMBH mass, providing a smooth transition between the radio and quasar modes \citep{Steinborn2015}. Mechanical outflows and radiation are both incorporated in form of thermal feedback.

%--------------------------------------------
\subsection{Cluster redshift subsamples and combined sample}
\label{sec:sample}
%--------------------------------------------

The simulated regions are processed through the \tre\ standard pipeline, whose first step is represented by the identification of structures and substructures using the Amiga Halo Finder\footnote{https://popia.ft.uam.es/AHF} \citep{ahf}. The center of the identified objects is located at the position of the local maximum of the total density distribution. With a spherical overdensity approach, we define $R_{\Delta}$ as the radius of the sphere enclosing a density equal to $\Delta$ times the critical density of the universe at the considered redshift and $M_{\Delta}$ as the total mass contained within such radius. In our work, we focus on $\Delta=2500, 500, 200$, which, respectively, might roughly be associated with the cluster's core region, with the area typically covered by X-ray and SZ observations, and with the radius often referred to as the virial radius (although see footnote $^1$).

For the specific analysis of this work in each region and for each snapshot we consider the most massive cluster that does not contain any low-resolution particles within $R_{100}$. This condition is more stringent than the one applied in other works by The300 collaboration requiring the total mass associated with all low-resolution particles within $R_{200}$ to be less than 4 per thousand of the total mass of the cluster.  Furthermore, in our sample selection we impose $M_{2500}> 1.27 \times 10^{13}$ \hinv \msun\ to ensure that even the central region ($R<R_{2500}$) is represented by at least ten thousand particles.

The non-contaminated clusters (i.e., without any low-resolution particle) are extracted from ten different simulation snapshots selected at the redshifts: $z = 0.07$, $0.14$, $0.22$, $0.33$, $0.46$, $0.59$, $0.78$, $0.99$, $1.22$, $1.32$. Note that since we always select the most massive cluster in each region, the cluster subsamples at the highest redshifts will contain objects that are not necessarily the progenitors of those in the lowest redshift subsamples. Still, since the chosen redshifts are only separated by approximately 1 Gyr (the typical  crossing time of a cluster-sized halo at $z=0$), it is likely that the cluster subsamples of two consecutive redshifts are not completely independent. This does not present an issue for the analysis presented in Sect.~\ref{sec:res_m}, where each redshift subsample is independently analyzed, while it could affect the results of Sect.~\ref{sec:res_mz}, where we analyzed the combined sample formed by all clusters at all redshifts. For this, we verify that our results hold when we exclusively combine the subsamples corresponding only to $z=0.07$, $0.46$ and $1.32$ (see Appendix~\ref{app:verify}). 

\begin{figure*}[h!]
    \centering
    \includegraphics[width=0.49\textwidth]{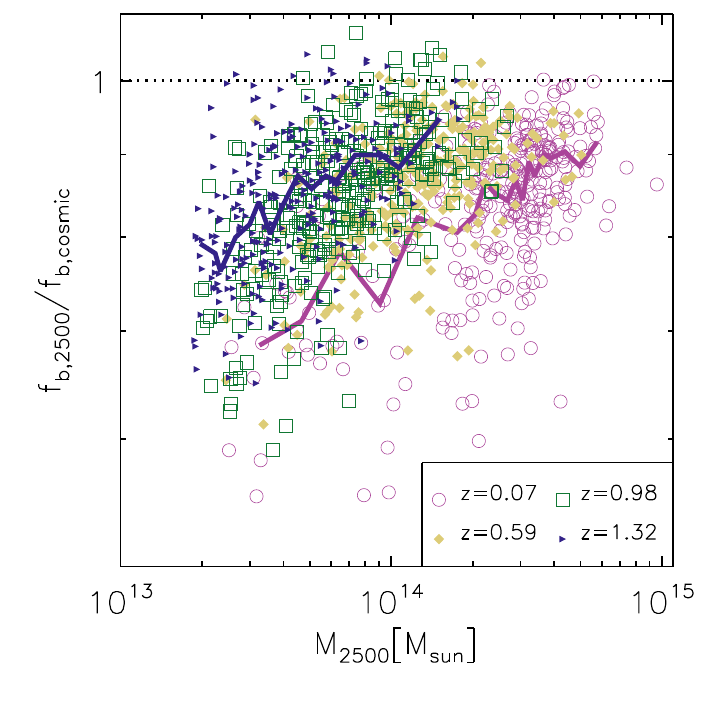}
   \includegraphics[width=0.49\textwidth]{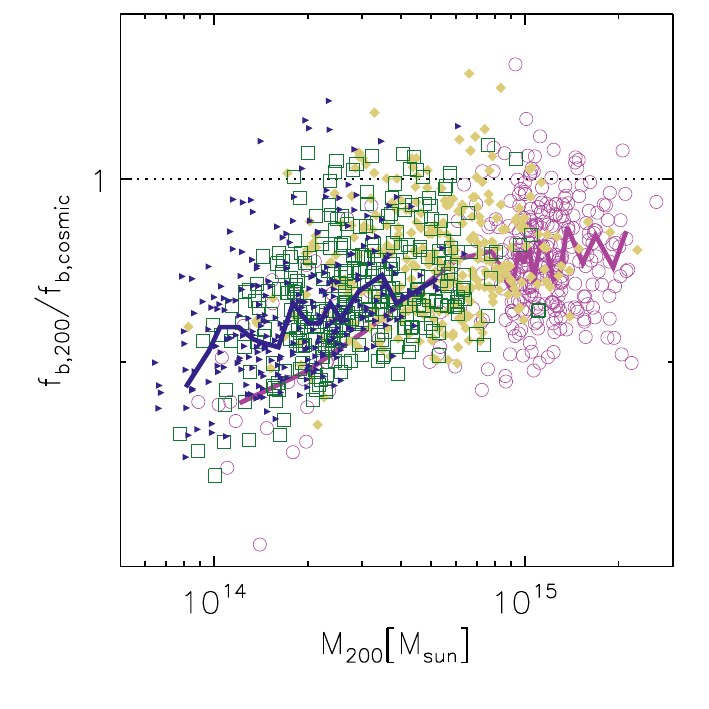}
\includegraphics[width=0.49\textwidth]{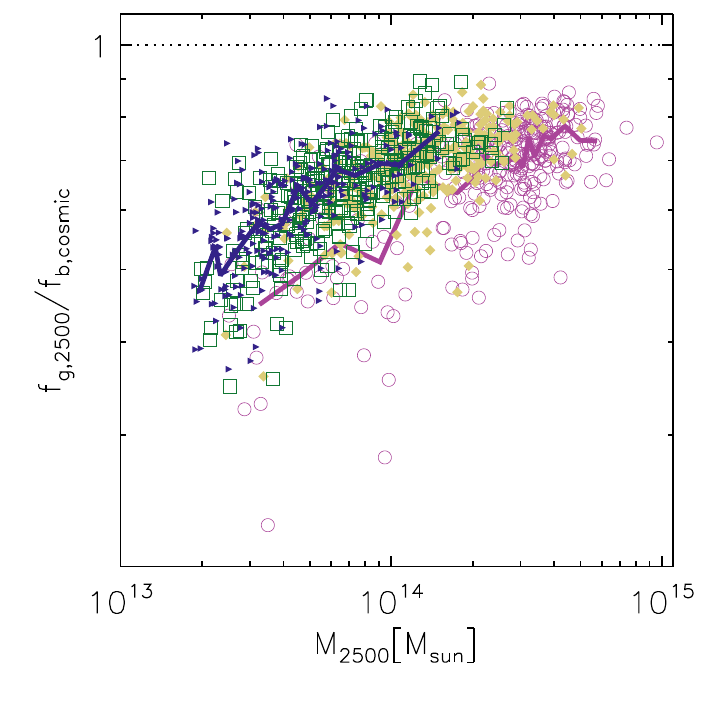}
   \includegraphics[width=0.49\textwidth]{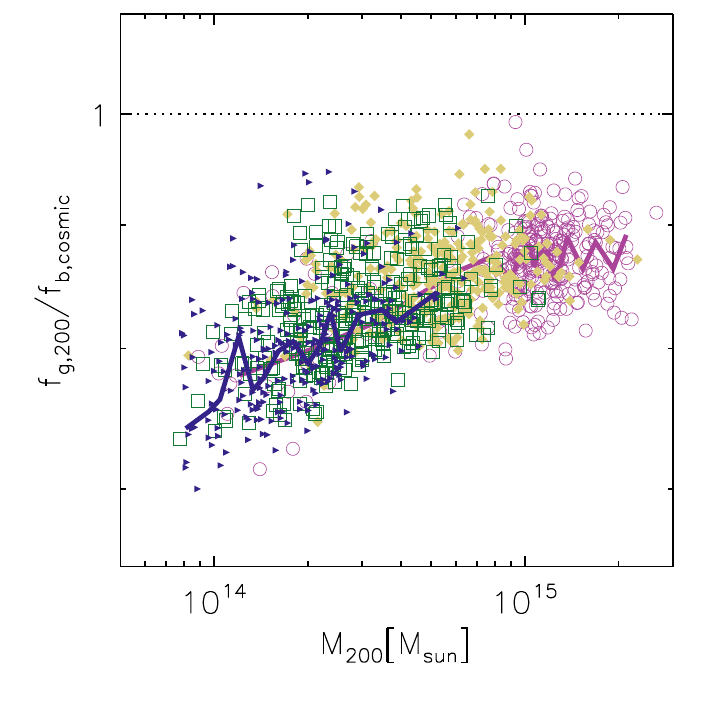}
    \caption{Baryon fraction (top panels) and hot-gas fraction (bottom panels) relative to the cosmic fraction versus the cluster mass within $\Delta=2500$ (left panels) and $\Delta=200$ (right panels). Data for $z=0.07$ (pink circles), $z=0.59$ (yellow diamonds), $z=0.98$ (green squares) and $z=1.32$ (navy triangles) are shown. The median values of the $z=0.07$ and $z=1.32$ subsamples are shown with a solid line. For reference: in both panels the maximum value of the y-axis is $1.1$ and  the horizontal line marks the cosmic baryon fraction.}
    \label{fig:barpresentation}
\end{figure*}

In Table~\ref{tab:sample}, we report the number of objects in each redshift subsample and their median mass at the three overdensities. 
The mass dependence of the baryon fraction and the hot-gas fraction, relative to the cosmic value, for the samples at $z=0.07$, $z=0.59$, $z=0.98$, and $z=1.32$ is presented in Fig.~\ref{fig:barpresentation} for the overdensities of $\Delta=2500$ and $200$. 
To compute the baryon fraction we consider both stellar and gas particles, while for the hot-gas fraction we limit the analysis to gas particles that are not forming stars and that have a temperature $T>0.3$ keV. The last condition is imposed to include only the gas particles that are expected to contribute significantly to emission in the soft X-ray band which provides the observational measurements of gas mass. We verify that the exact value of this threshold does not influence the results. In fact, the relative variation in the gas mass by decreasing this threshold to $T> 0.1$ keV is always less than 1.5 per thousand and less than $10^{10}$M$_{\odot}$ in absolute terms.

The figure will be discussed in detail later, but we note here some of the main aspects: the relative baryon fraction within $R_{200}$ is constrained between 0.9 and 1 for all objects and at all redshifts, whereas within the core regions it reaches much lower values. The fractions $f_{b,200}$ and $f_{g,200}$ have a mass dependence for masses $M_{200} < 3 \times 10^{14}$M$_{\odot}$, otherwise they are almost constant. The fractions $f_{b,2500}$ and $f_{g,2500}$, instead, are characterized by a large dispersion at all masses and redshifts. These results are due to the strong effect that non-radiative processes have on smaller radii and smaller masses.
%--------------------------------------------

\begin{table}
\caption{Basic properties the cluster samples.}   \label{tab:sample}      
\centering                         
\begin{tabular}{| c | c | c | c | c |}      
\hline
$z$ & $N$ & $M_{2500}$ & $M_{500}$ & $M_{200}$ \\  
&& [$10^{14}$ \msun]& [$10^{14}$ \msun]& [$10^{14}$ \msun]\\
\hline
0.07 & 322 & 2.81&  7.57 & 10.93 \\
0.14 & 321 & 2.46&  6.84 & 10.05 \\
0.22 & 323 & 2.08&  6.04 & 9.10 \\
0.33 & 323 & 1.82&  5.08 & 7.57 \\
0.46 & 323 & 1.51&  4.20 & 6.43 \\
0.59 & 322 & 1.25&  3.58 & 5.37 \\
0.78 & 323 & 0.94&  2.58 & 3.86 \\
0.99 & 318 & 0.67&  1.91 & 2.81 \\
1.22 & 292 & 0.50&  1.45 & 2.25 \\
1.32 & 275 & 0.43&  1.32 & 1.97 \\
\hline                        
\end{tabular}
\tablefoot{For each redshift (Col.~1), we report the number of objects satisfying both conditions on the absence of low-resolution particles and on the $M_{2500}$ minimum mass (Col.~2), and their median masses in units of $10^{14}$ \msun\ for the three considered overdensities, $\Delta=2500, 500, 200$ (Cols. 3--5).}
\end{table}

%--------------------------------------------

%--------------------------------------------
%--------------------------------------------
\section{Method of analysis}
\label{sec:met}
%--------------------------------------------

In this work, we consider various functional forms for the analytic description of the mass dependence of the baryon and hot-gas fractions. Expressing both variables in a logarithmic scale, we compare: the usual linear relation, and the parabolic and logarithmic relations similar to those proposed by \cite{stanek.etal.2010} and \cite{vikhlinin.etal.2009}, respectively. To include the dependence on redshift, we investigate three different strategies and further study whether this dependence is better characterized as a function of $(1+z)$ or as a function of $E(z)^2=[h(z)/h(0)]^2=\Omega_M \times (1+z)^3+\Omega_{\Lambda}$.  A comparison between the trends of $(1+z)$ and $E(z)$ is shown in Fig.~\ref{fig:evolratio}. 
 The fitting formulae that we consider in this work are introduced below, followed by a description of the adopted methodology.

%--------------------------------------------
\subsection{Functional forms}
%--------------------------------------------
\subsubsection{Dependence on the total mass}
\label{sec:mass}
%--------------------------------------------
As anticipated in the Introduction, a positive correlation between the baryon and gas fractions and the total cluster mass is found both in observations and in simulations. In fact, observations show evidence that the smallest objects (poor groups and galaxies) are depleted of baryons as a consequence of the intense AGN activity experienced at higher redshift, when a shallower gravitational potential well is unable to retain gas heated by feedback processes \citep{eckert.etal.2021}.
Following this reasoning, at each redshift we describe the relationship between baryon/gas fraction and mass with three different functional forms: 
\begin{itemize}
\item[$\bullet$] A linear fit:
\begin{equation}
Y=a_2+b_2\, X;
\label{eq:glob}
\end{equation}
\item[$\bullet$] A parabola, as originally proposed by \cite{stanek.etal.2010}:
\begin{equation}
Y=a_{\mathrm 3}+b_{3}\,  X +c_{3}\, X^2;
\label{eq:stanek}
\end{equation}
\item[$\bullet$] A logarithmic relation, as proposed by \cite{vikhlinin.etal.2009} but in the logarithmic plane:
\begin{equation}
Y=a_{4}+b_{4} \log(X+1.20).
\label{eq:log}
\end{equation}
\end{itemize}
In these relations, we define $X=\log(M_\Delta/M_0)$ with $M_0=3\times 10^{14}$\msun\ at all redshifts, and $Y= Y_{\Delta}=\log(f_{\Delta}/f_{\mathrm{b, cosmic}})$, where $f_{\Delta}$ is either the baryon or the hot-gas fraction. Throughout the paper, we will refer to this quantity, which is the complement to $1$ of the depletion factor, as the {\it relative fraction} and the pedices of the parameters will recall the number of the corresponding equation. A more extended description of the functional forms is presented in Appendix~\ref{app:bar}.

\subsubsection{Dependence on total mass and redshift}
\label{sec:red}
%--------------------------------------------
As a second step, we investigate how the previous relations vary with redshift. Therefore, after looking at the evolution of the best-fit parameters of Eqs.~\ref{eq:glob}, \ref{eq:stanek}, and \ref{eq:log} obtained in each redshift subsample (see Sect.~\ref{sec:res_m} and Appendix~\ref{app:bar}), we incorporate the redshift dependence in the expressions to fit the combined sample. Namely, we consider the following three alternatives (see Sect.~\ref{sec:res_mz}):
\begin{itemize}
\item[$\bullet$] An extension of the previous functions with a translation along the $Y$ axis:
\begin{equation}
Y=f(X)+d_{5} Z.
\label{eq:exty}
\end{equation}
\item[$\bullet$]An extension of the previous functions with a translation along the $X$ axis:
\begin{equation}
Y=f(X+d_{6} Z). 
\label{eq:extx}
\end{equation}
\item[$\bullet$] The redshift dependence as in \cite{planelles.etal.2013}
\begin{equation}
Y=a_7+b_7  X +\log(1+d_7 z). 
\label{eq:Zpla}
\end{equation}
\end{itemize}
In all these cases $X$ and $Y$ have the same meaning as before, $f(X)$ has to be intended as one of the three functional forms introduced in the previous subsection (the line, the parabola, and the logarithm in the logarithmic plane), $Z$ provides the functional form of the redshift dependence, and $d_5$, $d_6$, $a_7$, $b_7$ and $d_7$ are fitting parameters, whose subscripts refer to the number of the corresponding equations. For $Z$, we consider two expressions: $Z=\log{[({1+z})/({1+z_0})]}$ and $Z=\log[E(z)]$, where $z_0=0.53$, which is in the middle of the redshift range considered. 

In the above formulae, the redshift dependence corresponds to a family of curves with the same shape but translated along the $Y$ axis (Eq.~\ref{eq:exty}) or along the $X$ axis (Eq.~\ref{eq:extx}). While the first case represents a simple change in normalization or a shift in baryon fraction at fixed mass, the second expression implies that the quantity that changes with redshift is the total mass.

%--------------------------------------------
\subsection{Fitting procedure}
\label{sec:procedure}
%--------------------------------------------

Each of the redshift sample, used to study the dependence of the baryon and hot-gas fractions on the total mass, includes a number of objects varying between 275 and 324 (see Table~\ref{tab:sample}). Their combination provides a sample of $\approx 3150$ objects. 
To study either the redshift subsamples or the combined sample, we proceed by binning the data. Specifically,  we sort the clusters according to their mass (either $M_{2500}$, $M_{500}$, or $M_{200}$) and then we build mass bins starting from the most massive cluster. The bins with more sparse data (the first two more massive and the last 3-4 less massive) contain 10 clusters each, while all others combine 20 clusters. In this way the size of the mass bins are not too uneven. The least-massive bin has 10 objects plus any remaining one. We tested different binning choices and they proved to lead to statistically similar results.

To each mass bin $j$, we associate the medians of the $X_{i,j}$ and $Y_{i,j}$ values,  respectively $\mu_j$ and $\eta_j$,with $i$ representing the clusters belonging to that mass bin. We then compute their errors defined as the median absolute deviations and given, respectively, by
\begin{equation}
\label{eq:err_mu}
\sigma_{\mu,j}={\rm median}(|X_{i,j}-\mu_j|)
\end{equation}
and 
\begin{equation}
\label{eq:err_eta}
\sigma_{\eta_j}={\rm median}(|Y_{i,j}-\eta_j|).
\end{equation}

The fit procedure uses the {\sc IDL} routine {\sc mpfitfun} and is performed 1000 times with the X values allowed to randomly vary in the range $[\mu-\sigma_\mu,\mu+\sigma_\mu]$. The best-fit parameters and their errors are computed as the mean and standard deviation of the distribution of the obtained 1000 values. 
We verified that this approach, based on the Levenberg-Marquardt least-square fitting, provides a statistically similar result of the outlier-resistant two-variable linear regression of the routine {\sc robust\_linefit}.

To evaluate whether the analyzed models are a good description of the baryon and hot-gas fraction relations with the total mass, we compute the residuals expressed as a percentage:
\begin{equation}
\mathcal{R}=100 \times \frac{10^{\eta_{j}}-10^{Y_{\mathrm fit,j}}}{10^{\eta_{j}}}
\end{equation}

All the numerical values of the parameters and of their errors %, and of $\chi$ 
are reported in Appendix~\ref{app:bar} for the total baryon fraction and Appendix~\ref{app:gas} for the hot-gas fraction. There, we also report the values of the relative baryon fraction measured at $M_0=3\times 10^{14}$ \msun, from which, one easily obtains the value of the depletion factor at that mass scale.  
Since all values of the parameters are small numbers, in the tables and throughout the paper, we will report values that are 100 times larger than the parameters themselves and we will use capital letters to denote them, such as $A_2=100 \times a_2$.

%--------------------------------------------
\section{Comparison to observations}
\label{sec:obse}
%--------------------------------------------
Before exploring the dependence of the baryon and hot-gas fraction on the total cluster mass and its evolution, we compare the results from our simulations with some observational measurements \citep[see also][]{cui.etal.2018, cui.etal.2022}.  In Figs.~\ref{fig:fig2},  we plot, respectively, with black and navy dots, the simulated clusters of the subsamples at $z=0.07$ and $z=1.32$.

\subsection{Baryon fraction}

\begin{figure*}
\includegraphics[width=0.49\textwidth]{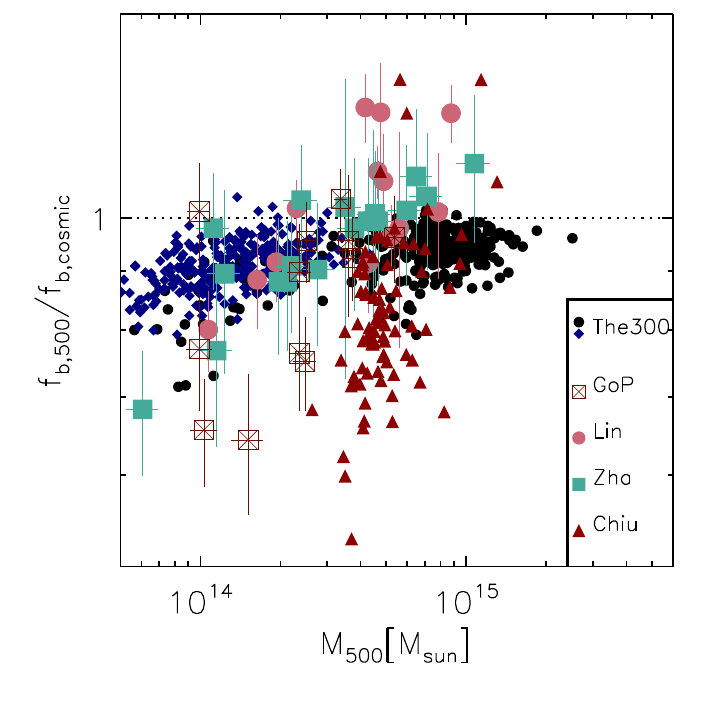}
\includegraphics[width=0.49\textwidth]{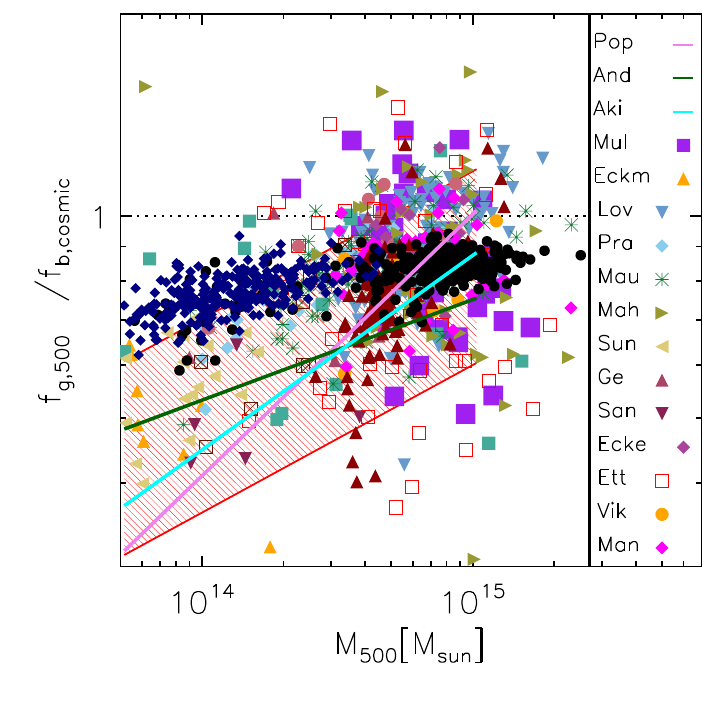}
    \caption{Baryon fraction (left panel) and gas fraction (right panel) relative to the cosmic fraction versus the cluster mass within $R_{500}$ at $z=0.07$ (black dots) and $z=1.32$ (navy dots). The legend associates colors and symbols to the shortenings of the observational papers of reference that include both local and distant clusters: 
Zha for \cite{zhang.etal.2011SF},
GoP for \cite{gonzalez.etal.2013},
Lin for \cite{lin.etal.2003},
Chiu for \cite{chiu.etal.2018},
Mul for \cite{mulroy.etal.2019},
Eckm for \cite{eckmiller.etal.2011},
Lov for \cite{Lovisari.etal.2020},
Pra for \cite{pratt.etal.2009},
Mau for \cite{maughan.etal.2008},
Mah for \cite{mahdavi.etal.2013},
Sun for \cite{sun.etal.2009},
Ge for \cite{ge.etal.2018},
San for \cite{sanderson.etal.2013},
Ecke for \cite{eckert.etal.2019},
Ett for \cite{ettori.etal.2009},
Vik for \cite{vikhlinin.etal.2006},
Man for \cite{mantz.etal.2016b}. 
The red band is from \cite{eckert.etal.2021}, and the purple, dark green, and cyan lines are, respectively, from \cite{popesso.etal.2024arxiv}, \cite{andreon.etal.2017}, and \cite{akino.etal.2022}.}
 \label{fig:fig2}
\end{figure*}
The baryonic fraction of the simulated clusters is shown on the left panel of Fig.~\ref{fig:fig2} and it is compared with the observational data described in the following, starting from the most recent one.

\cite{chiu.etal.2018} studied the baryon content of 91 clusters detected from the Sunyaev-Zeldovich effects by the South-Pole Telescope, and followed-up with the {\em Chandra} satellite in the X-rays. The sample includes only massive systems with $M_{500}> 2.5 \times 10^{14}$M$_{\odot}$ and extends from $z=0.2$ to $z=1.25$. One of the results of that paper is that the baryon fraction in their sample does not change with redshift and thus we include all their objects in Fig.~\ref{fig:fig2}. We do not apply any correction to either the total mass or the baryon fraction since the work assumed a value for the Hubble constant similar to the one considered in The300 simulation.

\cite{gonzalez.etal.2013} investigated the total baryon fraction in a sample of 12 local clusters with masses in the range $(1$--$5) \times 10^{14} M_{\odot}$.
For consistency with the cosmological model assumed in our simulations, we used the values of the baryon fraction reported in their table~6 which refers to the Planck cosmology. Notice that the paper by \cite{gonzalez.etal.2013} often serves as a reference for the stellar fraction-mass relation. Indeed, many subsequent papers that investigated the total baryon fraction are based on new measurements of only the gas fraction.

\cite{lagana.etal.2011} used a subset of the \cite{zhang.etal.2011} clusters extracted from the HIghest X-ray FLUx Galaxy Cluster Sample (HIFLUGCS). The gas mass was derived from XMM-Newton observations, while the stellar masses from the optical data of SDSS-DR7 \citep{abazajian.etal.2009}. The assumed  cosmology was from WMAP7 with the reduced Hubble parameter $h=0.7$, thus slightly higher than ours. To compare with this data, we appropriately re-scale their total mass in solar-mass units by $h_{\rm obs}/h_{\rm sim}$ and the gas fractions by $(h_{\rm obs}/h_{\rm sim})^{3/2}$. The correction for the stellar fraction and consequently for the baryon fraction is not trivial because different factors enter in play at different stages of the analysis. We then evaluate a correcting factor by comparing the tables corresponding to the cosmologies of WMAP7  and Planck presented in \cite{gonzalez.etal.2013}, and obtain $(h_{\rm obs}/h_{\rm sim})^{5/6}$

Finally, we consider the baryon fraction values obtained by \cite{lin.etal.2003}. Three of their clusters (A85, A1367, A2255) are in common with \cite{lagana.etal.2011} and the baryon fraction values are consistent with the exception of A85, for which the former paper provided a 10 percent higher value. The three objects with the largest relative baryon fraction (the three brown triangles which in the left panel of Fig.~\ref{fig:fig2} have relative baryon fraction larger than $1.1$) are A3558, A3266, and A2319: the first and the second clusters are part of the Shapley and Horologium superclusters, while the third is a well-known major merger system  and it is the only object in the X-COP sample with gas fraction in excess of the cosmic value \citep{ghirardini.etal.2018}. Possible presence of clumpiness or substructures and irregular geometry can bias the values of the total mass of these three objects and therefore explains the high value of the reported baryon fraction.
Also for this sample, we apply the above corrections due to the Hubble parameter difference.

\vspace{3pt}
Looking at the left panel of Fig.\ref{fig:fig2}, we consider the comparison between simulations and observations to be satisfactory at all mass scales and without any particular bias leading to an under-estimate or an over-estimate of the baryon fraction in our numerical objects. It is, however, noticeable that the scatter of the observed data is significantly larger than for numerical predictions. Part of this difference could be due to measurement biases, 
for example, part of the diffuse stellar component could be undetected and the gas masses  and total masses could be extrapolated at $R_{500}$, especially for the least massive clusters. At the same time,  sub-grid models in simulations are (by definition) simplistic approximations to very complex astrophysics processes covering more than ten orders of magnitude in scale, thus it is likely that the current implementation of the AGN feedback is unable to capture the variety of phenomena occurring in observed galaxy clusters, which strongly impact star formation, the onset of galactic outflows, and the overall distribution of the gas and the diffuse stellar components \citep{Oppenheimer.etal.21}. Some limitations can be associated to the pure thermal injection of the feedback that expands spherically (rather than with turbulent bubbles or cocoon shocks) whose energy is deposited at a larger-than expected radius \citep{hlavacek-larrondo.etal.2022}.  In addition, in these simulations high-redshift ($\gtrsim 2$) galaxies do not experience a phase of bursting star formation \citep{bassini.etal.2020}, therefore they do not have a  consequent substantial stellar feedback that also impacts the baryon distribution. As we will see in Sect.~\ref{sec:compasim}, the main features of our simulated clusters are common to almost all other simulated large datasets of clusters.

\subsection{Hot-gas fraction}
In the right panel of Fig.~\ref{fig:fig2},  we plot the relative gas fraction computed within $R_{500}$ at $z=0.07$  and $z=1.32$ as a function of the total cluster mass. 
As done previously, from \cite{gonzalez.etal.2013} we chose the values that refer to the Planck cosmology and, similarly, we consider the values by \cite{vikhlinin.etal.2006} revisited by \cite{ettori.etal.2009} to bring them to the cosmology with $h=0.7$ and $\Omega_M=0.3$. All observational values of total mass and gas fractions are then transformed via the Hubble parameter assumed in our simulations. In addition to individual cluster measurements, we add in the plot the shaded region that was identified by \cite{eckert.etal.2021} as to comprise other observational samples and the scaling relations provided by \citep{andreon.etal.2017,akino.etal.2022,popesso.etal.2024arxiv}. 

The observational data points contains both local and distant  objects (including $z>1$ systems).
%, but the gas fraction 
% of massive systems ($M_{500}>2\times 10^{14}$M$_{\odot}$) within $R_{500}$ is not expected to significantly change in the redshift range considered (see discussion below in Sect.~\ref{sec:massive}). 
 We note an overall agreement between our results and the observations for systems with mass $M_{500} > 2 \times 10^{14}$M$_{\odot}$.  For the less massive clusters, there is an indication that our simulated objects are gas richer than the observations (e.g., \citealt{popesso.etal.2024arxiv}) even though some observational data are still in agreement with the bulk of the simulated results. In general, the possible higher gas fraction in groups in our simulations could lead to a weaker mass dependence than in observations, with a normalization matching the observed one in the high-mass end. If anything, then, the transition between clusters and small groups could be more prominent in the observed samples than in The300. We stress that the comparison by \citet{cui.etal.2022} showed that in the group mass regime the stellar fraction is slightly overestimated (see discussion in Sect.~\ref{sec:discussion}) so that the combined contribution of hot gas and stars leads to an overall baryon budget in agreement with observations, as shown in the left panel of Fig.~\ref{fig:fig2}.

As noted in the previous section, the most apparent difference between observations and simulations is the scatter of the data, also for the hot gas fraction. For example, the observational samples of \cite{Lovisari.etal.2020} and \cite{maughan.etal.2008}, which respectively contain 120 and 115 objects, have a spread around the median about 3 times larger than that of The300 and the smaller samples of \cite{gonzalez.etal.2013}, \cite{zhang.etal.2011SF}, and \cite{mulroy.etal.2019}, which contains about 15-40 objects, have a scatter as much as 6 times larger than the simulated one.
There are several possible explanations for this difference. Besides the causes mentioned above, for the hot-gas component is important to notice that at the low temperature of galaxy groups, the cooling function is highly sensitive to the assumed metallicity, complicating the derivation of the gas density from surface brightness profiles. 
Indeed, measurements of the gas fraction are typically associated to large error bars, while simulated data are not, thus a fraction of the observed scatter is also due to statistical uncertainties. 
In addition, in Fig.~\ref{fig:fig2} the various observational works differ for sample selection, and methodologies of analysis. In particular, some analyses are based on data from XMM-Newton, others from Chandra, from a combination of the two satellites, or even from a combination of X-ray and SZ data. Some samples are X-ray selected, others are SZ or optically selected. To measure the total mass, some of the analyses assume the hydrostatic-equilibrium, which can be affected by several biases \citep{rasia.etal.2012,biffi.etal.2016,ansarifard.etal.2020, gianfagna.etal.2023}, others assume a scaling relation between total mass and temperature or galaxy velocity dispersion. Other analyses are based on masses from gravitational lensing, which can be affected by projection effects \citep[e.g.][and references therein]{giocoli.etal.2023}. All these factors contribute to the increase of the spread among the observational samples. Indeed objects, common among different samples, have values of total mass and gas fraction that are discrepant at a high significance. One example is A2029 which appears in four observational samples\footnote{The mass and gas fraction values reported here refer to the cosmology used in this paper.}, with a total mass ranging from $8.7$ \citep{mantz.etal.2016b} to $15.2 \times 10^{14}$M$_{\odot}$ \citep{zhang.etal.2011SF} and gas fraction from 0.095 \citep{zhang.etal.2011SF} to 0.148 \citep{eckert.etal.2019}. Moreover, the relative gas fraction changes from 0.94 in \cite{eckert.etal.2019} to 0.86 in \cite{vikhlinin.etal.2006} despite the two papers report a similar total mass of about $9\times 10^{14}$M$_{\odot}$. This said, and as noted above, the larger scatter displayed by observational results could also tell us something about the inability of the sub-resolution models of star formation and stellar and AGN feedback to capture the complexity of the baryon cycle at the scale of groups and clusters. We stress that these possible limitations are expected to have more impact on the hot-gas distribution than on the total baryon content and to affect more deeply the smallest mass systems \citep{davies.etal.2020,donahue_voit}.

%--------------------------------------------
\section{Baryon fraction: results on redshift subsamples}
\label{sec:res_m}
%--------------------------------------------

The detailed results of the fitting procedure of the relation between the total mass and the baryon fraction are described in Appendix~\ref{app:bar}, where we also show in the figures and list in the tables all best-fit parameters for all redshift subsamples. In Appendix~\ref{app:medbar}, we present the evolution of the medians of all clusters in each subsample and show that there is little evolution across the redshift range here investigated.

\vspace{3pt}

The main results of the analysis presented in Appendix~\ref{app:bar} for the three functional forms can be summarize as follows: 
\paragraph{Linear form:} the best-fit normalizations and slopes of the linear form are approximately constant with redshift for both $R_{200}$ and $R_{500}$, while the normalization at $R_{2500}$ slightly grows at high redshift (see Eq.~\ref{eq:evnorm}); 
\paragraph{Parabolic form:} the best-fitting parameters of the parabolic form do not show any evolution, in addition the values of the curvature parameter for $R_{200}$, almost always, is equal to zero underlining that the relation between the baryon fraction and the total mass at low overdensity can be fairly represented with a simple power law, while this statement is not true at higher overdensities;
\paragraph{Logarithmic form:} a small trend with redshift can be found in both the best-fitting normalizations and slopes of the logarithmic form (see Eq.~\ref{eq:evnormlog} and Eq.~\ref{eq:evslopelog}) at $R_{2500}$. The logarithmic curvature parameter, $b_4$, is never exactly equal to zero (see Table~\ref{tab:eq3}), implying that a constant baryon fraction is never the best-fitting solution within the redshift and mass range considered.

%-----------------------------------

\begin{figure}
    \centering
\includegraphics[width=0.49\textwidth]{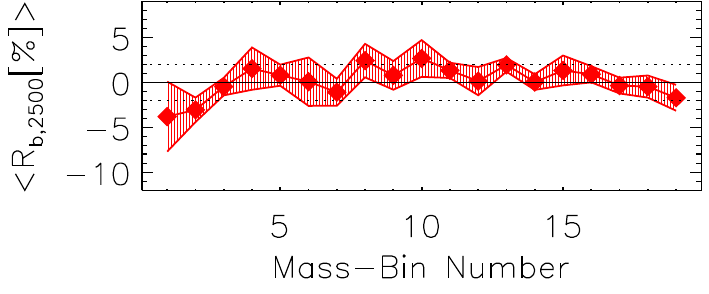}
\includegraphics[width=0.49\textwidth]{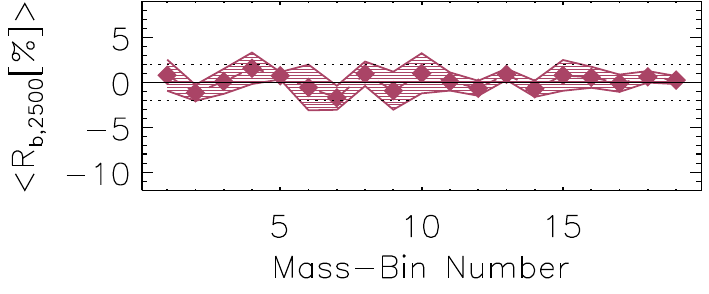}
\includegraphics[width=0.49\textwidth]{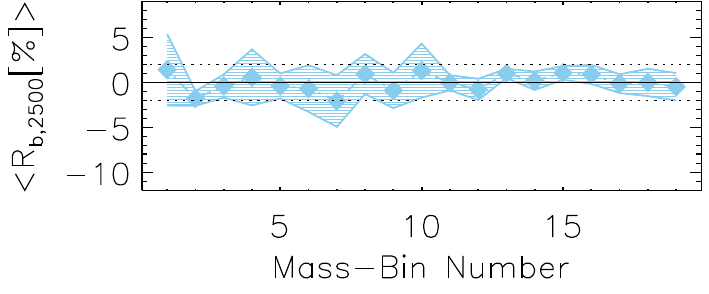}
    \caption{Averaged residuals at $R_{2500}$ of the fitting function in Eq.~\ref{eq:glob} (top panel), Eq.~\ref{eq:stanek} (middle panel), and Eq.~\ref{eq:log} (bottom panel). The shaded area represent 1$\sigma$ deviation from the mean.
    }
    \label{fig:newsec5}
\end{figure}

In Fig.~\ref{fig:newsec5} the averages of the residuals of the fitting functions are shown.   
Since the mass range, thus the number of bins, is different in the various subsamples, the average is computed after aligning the first bin, corresponding to the smallest masses. 
The residuals of the three models are of comparable order, however, the largest deviations are related to the power-law formula (top panel) and they are all associated to a worse performance of the fit for the least or the most massive bins. 

\begin{figure*}
    \centering
   \includegraphics[width=0.95\textwidth]{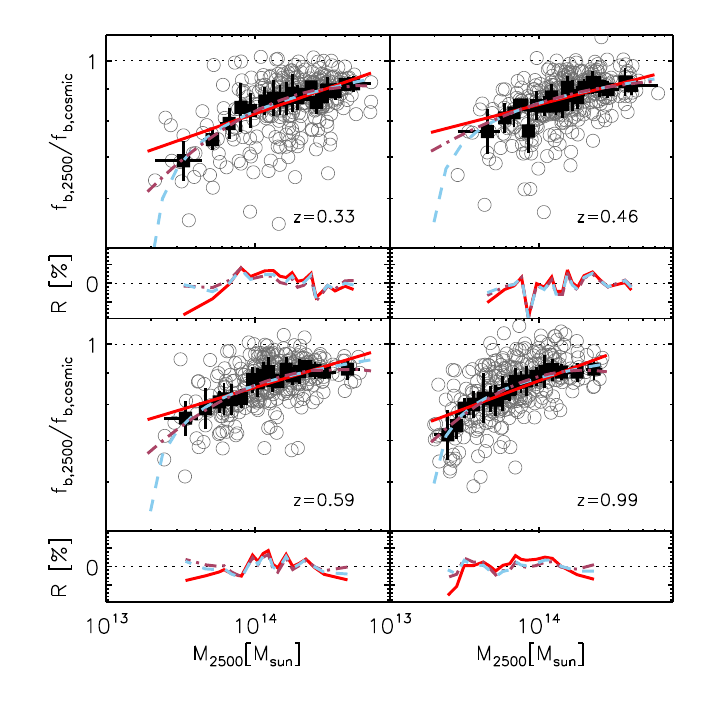}
    \caption{Best-fit functions of Eq.~\ref{eq:glob} (red solid line), Eq.~\ref{eq:stanek} (brown dot-dashed line), and Eq.~\ref{eq:log} (light-blue dashed line) at the four redshifts specified in each panel. Individual clusters are shown with gray empty circles, while the median values are shown with black squares. Error bars indicate $\sigma_\mu$ and $\sigma_\eta$. The residuals of the best-fit functions with respect to the median values are shown at the bottom of each panel.}
    \label{fig6}
\end{figure*}

Another visualization of this effect is presented in Fig.~\ref{fig6}, where we show the baryon fraction within $R_{2500}$ at $z=0.33, 0.46, 0.59$ and $z=0.99$, the corresponding best-fitting functions, and their residuals. 
All residuals have visible fluctuations, due to the scatter among the fraction medians, but those related to the linear fit (in the logarithmic plane) are `systematically' off in the first few and last bins, indicating an overestimation of the data by the linear fit.
The slight improvement obtained by using the quadratic or the logarithmic formulae is due to the fact that both fitting functions appropriately capture the flattening of the mass-dependence of the baryon fractions at the highest masses and its steepening as the cluster mass decreases. 
 The extrapolations of the two functions to the lowest value of $M_{2500}$ highlight how they better describe not only the median values of the baryon fraction but also the distribution of the individual clusters. For completeness, we report that the residuals at the lowest overdensities tend to be always close to zero for all functional forms.

%--------------------------------------------
%--------------------------------------------
\section{Baryon fraction: results on the combined sample}
\label{sec:res_mz}
%--------------------------------------------
%--------------------------------------------

In observational studies binning the data in small redshift intervals, as done in the previous section, is not always achievable since it would require a large amount of data that densely populate the redshift intervals of interest. In this section, we then follow a more observational-oriented approach: we simultaneously fit all clusters together and consider the time dependence explicitly in the equations. In practice, the fits are applied to the median relative fractions for each mass bin and redshifts.

The fitting procedure performed on the redshift subsamples presented in the previous section has shown that the best-fit parameters have a regular behavior from $z=0$ to $z=1.3$. The normalization at $R_{2500}$ was the only parameter which showed some trend with redshift, albeit small. We therefore expect that the evolution might be well captured by the expression in Eq.~\ref{eq:exty} which represents a family of curves translated along the y-axis or, in other words, characterized by the same shape but different normalization. However, the theoretical expectation, when only gravitational processes are at play, is that the baryon fraction does not change with time while the total mass varies with $E(z)^{-1}$. This scenario would be better captured by Eq.~\ref{eq:extx}.
These two approaches are discussed in the following two subsections. We remind that in Appendix~\ref{app:verify}, we discuss the results from the same analysis applied to only three redshift subsamples ($z=0.07$, $z=0.46$, and $z=1.32$) which can be considered independent.

%---------------------
\subsection{Redshift dependence of the normalization}
\label{sec:exty}
%---------------

The median values of relative fractions and masses from all sub-samples are used to evaluate the explicit redshift dependence of the baryon fraction according to Eq.~\ref{eq:exty}, where we consider as $f(X)$ the previously analyzed expressions: the line, the parabola and the logarithm in the logarithmic plane. Similarly to the previous approach, we iterate 1000 times where the mass of the bin is randomly taken from an interval ([$\mu-\sigma_\mu/2:\mu+\sigma_mu/2$] in the case) and the error on the fractions are accounted in the fitting procedure. The mean and standard deviation of the 1000 iteration best-fit values of the parameters are reported in Table~\ref{tab:table2} for the three considered over-densities, and for the redshift dependence expressed as function of $(1+z)$. 
We have tested also the $E(z)$ parametrization, which is used in the self-similar model to re-scale virial quantities measured at different cosmic epochs \citep{kaiser86}. The retrieved parameters in this case are quite similar to the previous one and thus not reported in the table. The similar results for these two time-dependence expressions might be due to the fact that the differences between $E(z)$ and $(1+z)$ for the cosmology and the redshift range studied in this work are small (see Fig.~\ref{fig:evolratio}), especially compared to the dispersion of the data (see Fig.~\ref{fig:barpresentation}). The residuals related to the two expressions are also extremely close as can be seen by the minimal differences between the solid and dotted lines in the first panel of Fig.~\ref{fig:resb}, where we compare them for the linear fit. The differences for the other two fitting formulae are similar.

\begin{table}
\caption{Baryon fraction: best-fit parameters for Eq.~\ref{eq:glob}, Eq.~\ref{eq:stanek}, and Eq.~\ref{eq:log} with the $(1+z)$ redshift dependence as in Eq.~\ref{eq:exty}. }    
\label{tab:table2}      
\centering      
\begin{tabular}{|c |c c| c c| c c|}   
\hline
Eq.~\ref{eq:exty} & \multicolumn{2}{c|}{} & \multicolumn{2}{c|}{} & \multicolumn{2}{c|}{}\\
$f(X$)=Eq.~\ref{eq:glob} &\multicolumn{2}{c|}{$\Delta=2500$}  & \multicolumn{2}{c|}{$\Delta=500$} & \multicolumn{2}{c|}{$\Delta=200$} \\    
\hline 
$A_{2,z} \ \sigma_A$ & -2.8 & 0.3 & -3.3 & 0.1 & -3.0 &  0.1 \\
$B_{2,z} \  \sigma_B$ & 9.4 & 0.1 & 3.6  & 0.1 & 1.7  &  0.1  \\
$D_{2,z} \ \sigma_D$ & 21.1 & 0.7 & 4.8  & 0.2 & -0.1 &  0.1 \\
\hline
Eq.\ref{eq:exty} & \multicolumn{2}{c|}{} & \multicolumn{2}{c|}{} & \multicolumn{2}{c|}{}\\
$f(X)$=Eq.~\ref{eq:stanek}&\multicolumn{2}{c|}{$\Delta=2500$}  & \multicolumn{2}{c|}{$\Delta=500$} & \multicolumn{2}{c|}{$\Delta=200$} \\    
\hline   
$A_{3,z}\ \sigma_A$ &-3.0  & 0.2 & -2.9  &    0.1  & -2.9  &    0.1   \\
$B_{3,z}\ \sigma_B$ & 5.6  & 0.3 & 3.5  &    0.3  & 2.3  &    0.1 \\
$C_{3,z}\ \sigma_C$ &-4.0  & 0.1 & -2.4  &    0.3  &  -1.4  &    0.2 \\
$D_{3,z}\ \sigma_D$ &19.4  & 0.3 &  4.1  &    0.4  & -0.2  &    0.1 \\
\hline
Eq.~5 & \multicolumn{2}{c|}{} & \multicolumn{2}{c|}{} & \multicolumn{2}{c|}{}\\
$f(X$)=Eq.~\ref{eq:log} &\multicolumn{2}{c|}{$\Delta=2500$}  & \multicolumn{2}{c|}{$\Delta=500$} & \multicolumn{2}{c|}{$\Delta=200$} \\   
\hline 
$A_{4,z} \ \sigma_A$ & -4.9 & 0.1 & -4.1 & 0.2 & -3.4 &  0.1 \\
$B_{4,z} \  \sigma_B$ & 11.2  & 0.9 & 9.9  &0.6 & 5.6  & 0.1 \\
$D_{4,z} \ \sigma_D$ & 15.2 & 2.0 & 4.5  & 0.6 & 0.0 &  0.1 \\
\hline
%%%%%
\end{tabular}
\tablefoot{Best-fit parameters obtained from 1000 Montecarlo iteration using the {\sc idl} routine {\sc mpfitfun}. The parameters are multiplied by 100 (see comment at the end of Sect.~\ref{sec:met}).}
\end{table}

The results of the fitting procedure are shown in Fig.~\ref{fig7}, where we plot the median values of the relative baryon fractions associated to all sub-samples in gray, and we color those at the lowest redshift ($z=0.07$) in dark pink, at the intermediate redshift ($z=0.59$) in green, and at the highest redshift ($z=1.32$) in dark blue. We overplot the results of the fitting functions for Eq.~\ref{eq:exty} evaluated for the two extremes of the redshift range: $z=0.07$ and $z=1.32$. 
The overall evolution is more pronounced for the largest overdensity. Indeed, for all versions of the fitting function, the best-fit values of the parameter $D$, which is associated to the time evolution, at $\Delta=2500$ are larger by factors $\sim 3$-$4$ and $15$-$20$  with respect to $\Delta=500$ and $\Delta=200$. A significant spread among the redshift samples is still visible for $\Delta=500$, where the three colored sub-samples are still distinguishable. On the other hand, for $\Delta=200$ we do not detect any redshift dependence: in this case, the curve corresponding to $z=0$ is almost coincident with that of $z=1.32$; the dispersion of all points in the plot is driven by the spread within each redshift subsample; the values of the $D$ parameter in Table~\ref{tab:table2} are always consistent with zero within $1\sigma$.
Finally, we note that the mass corresponding to the peak of the parabola of $R_{200}$, $M_{200}\sim 1.3 \times 10^{15}$M$_{\odot}$, is within the mass range sampled by our set of simulated clusters, and thus the best-fitting function in principle includes also the descending part of the parabola. Nevertheless, the large curvature of the parabola makes the trend at high mass consistent to be flat. 

The average residuals of the three fitting function using Eq.~\ref{eq:exty} are shown in the first three panels of Fig.~\ref{fig:resb}, where we focus on the most relevant bins (the first and last few). In this case, we, indeed, show the average once the residuals are aligned according either to the first (left panels) and last (right panels) bin. Comparing these residuals with those of the redshift subsamples, we can see that imposing that the fitting parameters are the same throughout time worsens the performance in the least massive bins both for the linear form (top panel) and the logarithmic one (third panel), which are the analytical forms for which we note an evolution of either the slope or the curvature.

%-----------------------------
\subsection{Redshift dependence of the total mass}
\label{sec:extx}
%-----------------------------

\begin{figure}
    \centering
   \includegraphics[width=0.49\textwidth]{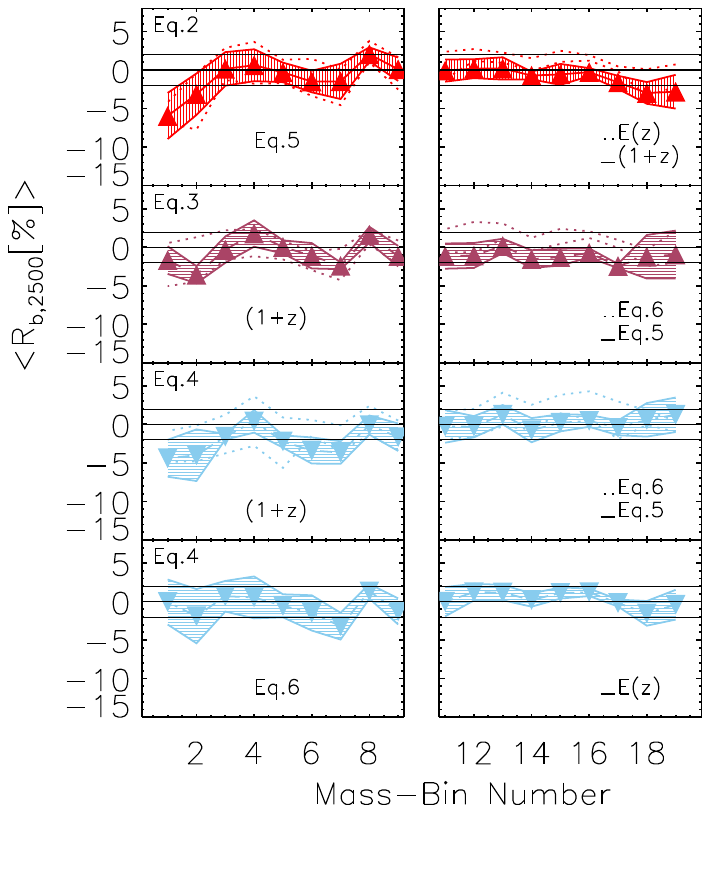}
    \caption{Average and 1$\sigma$ deviation of the residuals of the baryon fraction at $R_{2500}$ at fixed mass-bin number. Eq.~\ref{eq:glob}, Eq.~\ref{eq:stanek}, and Eq.~\ref{eq:log} are shown in the first, second, and third panel from the top, respectively, once the evolution is parametrized through $(1+z)$ as a shift in the normalization (Eq.~\ref{eq:exty}).  The dotted lines represent in the first panel the $E(z)$ evolution for the linear fit and in the second and third panels the residuals of Eq.~\ref{eq:extx}. The bottom panel shows the results of using Eq.~\ref{eq:log}, but after adopting $E(z)$ in the expression for the redshift dependence provided by Eq.~\ref{eq:extx}. Residuals have been aligned towards the first bin on the left panels and towards the last bin on the right panels.   }
    \label{fig:resb}
    \end{figure}

\begin{figure}
    \centering
  \includegraphics[width=0.5\textwidth]{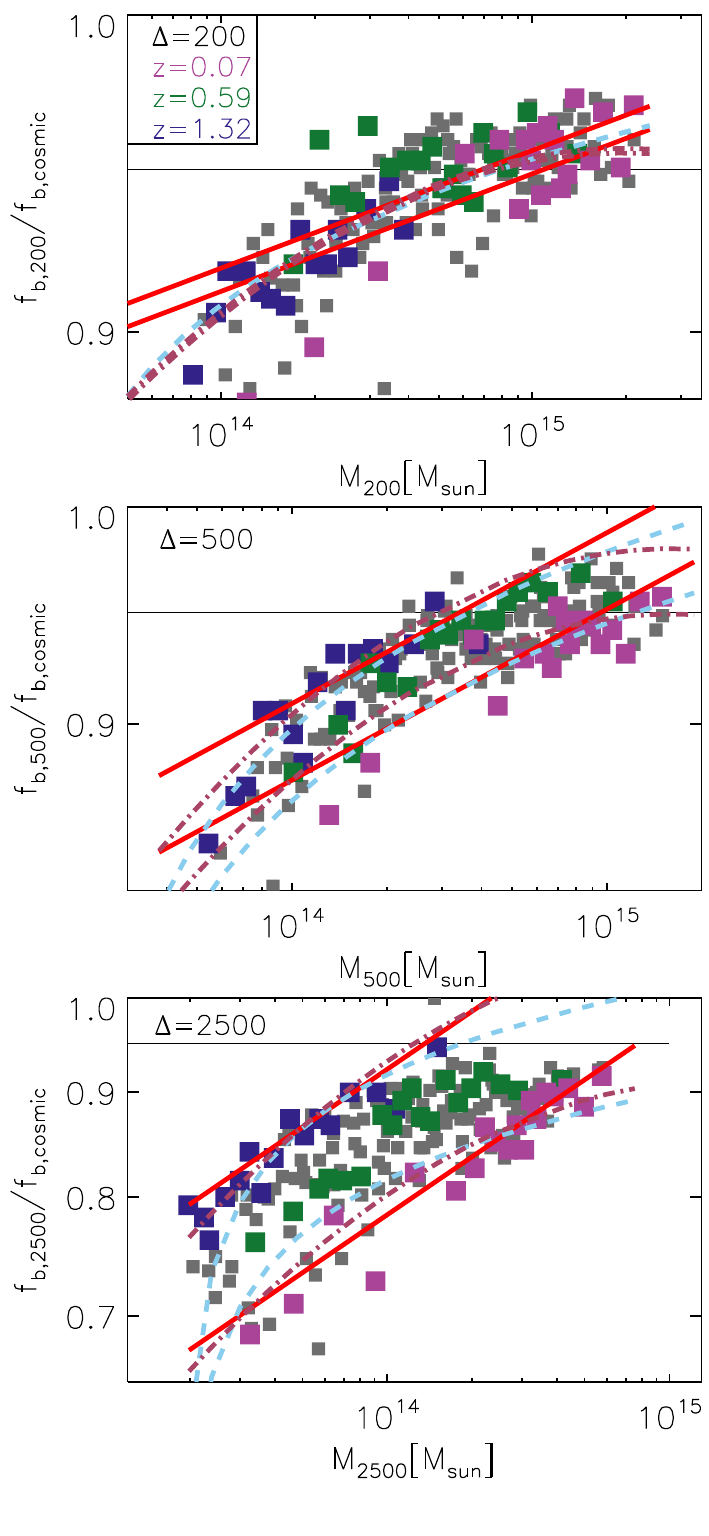}
    \caption{Median baryon fraction at all redshifts (plotted in gray) and at the three overdensities: $200$ (top panel), $500$ (central panel), and $2500$ (bottom panel). Subsamples at $z=0.07$, $z=0.59$, and $z=1.32$ are in dark pink, green, and  blue, respectively.
    Solid red, dot-dashed brown and dashed  light-blue lines represent the best-fit results using the $(1+z)$ dependence as in Eq.~\ref{eq:exty} for the three functional forms at $z=0$ and $z=1.32$ ( parameters in Table~\ref{tab:table2}). The horizontal line, equal to $0.95$, serves only as reference to appreciate the change in the y-axis range and it is  in all three panels.}
    \label{fig7}
    \end{figure}

Another possible redshift dependence representation is obtained by the family of curves that shift along the $X$ axis, i.e. the total cluster mass. As already mentioned, this picture is expected under the assumption that  gravity dominates the evolution of the baryonic content of clusters and groups. For obvious reasons, the results from Eq.~\ref{eq:exty} and Eq.~\ref{eq:extx} are identical if we consider the power-law form. For the other two expressions we report the best-fit parameters in Table~\ref{tab:z_o} where, at the bottom, we also add the best-fit results of Eq.~\ref{eq:Zpla}. 
With respect to the results discussed in the previous section, we notice that Eq.~\ref{eq:extx} leads to the same parameters for the baryon fraction within $R_{200}$ and $R_{500}$ due to the small redshift evolution that is registered at these overdensities (see first and second panel of Fig.~\ref{fig7}). At $R_{2500}$ there are some variations although the sets of parameters are consistent within 1$\sigma$. The second and third panel of Fig.~\ref{fig:resb} show that the residuals are also very similar (comparison between solid and dotted lines) even though Eq.~\ref{eq:extx} is associated to a larger dispersion in the first bins. In the logarithm case, the residuals of the first bins are overall improved when the $E(z)$ expression is used to parametrize the evolution within Eq.~\ref{eq:extx}. This combination, shown at the bottom panel of the figure, seems to perform better than either the vertical and the horizontal shift when $(1+z)$ is adopted (both curves in the third panel of the figure). Therefore, while for the vertical shift the difference between $(1+z)$ and $E(z)$ are negligible with respect to the dispersion of the data, they play a role when the shift is along the total mass which translates into a shifting in mass of the vertical asymptote. 

For the case of Eq.~\ref{eq:extx} and the $E(z)$ parametrization, we report in Appendix~\ref{app:dforez} the values of the $D$ parameter. Notice that for Eq.~\ref{eq:glob} (the linear fit) the best fit values are never favoring the self-similar evolution.

\begin{table}
\caption{Baryon fraction: best-fit parameters for Eq.~\ref{eq:glob}, Eq.~\ref{eq:stanek}, and Eq.~\ref{eq:log} with the $(1+z)$ redshift dependence as in Eq.~\ref{eq:extx}. In the bottom part: best-fit parameters for Eq.~\ref{eq:Zpla}.}      
\label{tab:z_o}      
\centering      
\begin{tabular}{|c |c c| c c| c c|}   
\hline
Eq.\ref{eq:extx} & \multicolumn{2}{c|}{} & \multicolumn{2}{c|}{} & \multicolumn{2}{c|}{}\\
$f$=Eq.~\ref{eq:glob}&\multicolumn{2}{c|}{$\Delta=2500$}  & \multicolumn{2}{c|}{$\Delta=500$} & \multicolumn{2}{c|}{$\Delta=200$} \\    
\hline   
$A_{2,z}\ \sigma_A$ & -4.4 &   1.0& -3.4  &    0.2  & -2.9   &   0.1   \\
$B_{2,z}\ \sigma_B$ & 7.4  &   1.5&  3.2 &    0.7  &  1.7   &   0.1\\
$D_{2,z}\ \sigma_D$ &203.5 &   15.0 &105.4  &   43.7  & -4.2  &   10.6\\
\hline
Eq.\ref{eq:extx} & \multicolumn{2}{c|}{} & \multicolumn{2}{c|}{} & \multicolumn{2}{c|}{}\\
$f$=Eq.~\ref{eq:stanek}&\multicolumn{2}{c|}{$\Delta=2500$}  & \multicolumn{2}{c|}{$\Delta=500$} & \multicolumn{2}{c|}{$\Delta=200$} \\    
\hline   
$A_{3,z}\ \sigma_A$ & -3.9 &   0.3& -3.0  &    0.1  & -2.9   &   0.1   \\
$B_{3,z}\ \sigma_B$ & 6.2  &   0.1&  3.6 &    0.3  &  2.3   &   0.1\\
$C_{3,z}\ \sigma_C$ & -4.5 &   0.3& -3.5  &    0.4  &  -1.4  &   0.1 \\
$D_{3,z}\ \sigma_D$ &210.2 &   2.1 &97.9  &   12.6  & -8.9  &   5.5\\
\hline
Eq.\ref{eq:extx} & \multicolumn{2}{c|}{} & \multicolumn{2}{c|}{} & \multicolumn{2}{c|}{}\\
$f$=Eq.~\ref{eq:log} &\multicolumn{2}{c|}{$\Delta=2500$}  & \multicolumn{2}{c|}{$\Delta=500$} & \multicolumn{2}{c|}{$\Delta=200$} \\    
\hline 
$A_{4,z} \ \sigma_A$ &  -4.9  &     0.1&  -4.3  &     0.1 & -3.3  &     0.1 \\
$B_{4,z} \  \sigma_B$&  10.6  &     1.1&  10.1  &     0.4 & 5.4   &     0.1 \\
$D_{4,z} \ \sigma_D$ & 31.8  &     43.2& 113.4  &    8.1  &-16.1 &    8.2 \\
\hline
\hline
%%%%%
& \multicolumn{2}{c|}{} & \multicolumn{2}{c|}{} & \multicolumn{2}{c|}{}\\
Eq.~\ref{eq:Zpla} &\multicolumn{2}{c|}{$\Delta=2500$}  & \multicolumn{2}{c|}{$\Delta=500$} & \multicolumn{2}{c|}{$\Delta=200$} \\    
\hline   
$A_{7,z} \ \sigma_A$ & -6.2 & 0.2 &-3.9 & 0.1  & -2.9  &    0.1 \\
$B_{7,z} \ \sigma_B$ &  9.2  & 0.1 & 3.5 & 0.1  &  1.7  &    0.2\\
$D_{7,z} \ \sigma_D$ & 14.2  & 0.5 & 2.7 & 0.1  &-0.2  &    0.1 \\
\hline 
\end{tabular}
\tablefoot{See notes of Table~\ref{tab:table2}.}
\end{table}

%--------------------------------------------

\subsection{Redshift dependence as in Planelles et al. (2013)}
\label{sec:plan}
%--------------------------------------------

In the bottom part of Table~\ref{tab:z_o} we add the best-fit parameters for Eq.~\ref{eq:Zpla}, which represents a redshift dependence previously presented in the literature. The results of this case have not been shown in any figure because the residuals associated with this description are significantly worse than those of the first two parts of the same table or those of Table~\ref{tab:table2}. This implies that the redshift dependence term as in \cite{planelles.etal.2013} is disfavored by The300 simulations in the mass and redshift ranges considered. 

%---------------
%---------------
\section{Hot gas fraction}
\label{sec:gas}

While the baryon fraction is a relevant quantity to measure in galaxy clusters for cosmological applications, the observational measurements tend to focus on the hot-gas fraction, that, we recall, in our simulations, is obtained by summing over all gas particles that are not star-forming and with temperature larger than $0.3$ keV.  
Similarly to the baryon fraction, the evolution of the medians of all clusters in each redshift bin is shown in Appendix~\ref{app:gasmedian}.

\subsection{Results on redshift subsamples}
The means and standard deviations of the 1000 iteration best-parameters of the redshift sub-samples fitting are discussed and reported in Appendix~\ref{app:gas} for Eq.~\ref{eq:glob}, Eq.\ref{eq:stanek} and Eq.\ref{eq:log}. 

The relation between the total mass and the hot-gas fraction in each redshift subsamples behaves very similarly to the baryon fraction, with two relevant differences: the normalization is lower, as expected, and the curvature parameters are typically larger, making the power-law description even poorer than for the baryon-fraction case. 
The more pronounced curvature is due to the fact that the ratio between stellar and gas content changes with mass and it is higher in smaller systems, which have then proportionally less hot-gas mass (see \citealt {angelinelli.etal.2023} and \citealt{biffi.etal.2025}). For example, the 20 least massive objects at $z=0.07$ have an average stellar-to-gas mass ratio at $R_{2500}$ which is almost twice as large than the same average ratio for the 20 most massive simulated clusters. Namely, $<M_{*}/M_{\mathrm gas}>=0.32$ for $M_{500}<4 \times 10^{14}$M$_{\odot}$, and  $<M_{*}/M_{\mathrm gas}>=0.17$ for $M_{500}>12 \times 10^{14}$M$_{\odot}$.  While the exact sharing of baryons between stars and cold or hot gas depends on the simulations sub-grid models, this trend is seen in all modern radiative simulations as discussed in Sect.~\ref{sec:compasim}.

\subsubsection{Massive clusters} 
\label{sec:massive}
We repeat the analysis presented above using only massive clusters, meaning with mass $M_{500}>2\times10^{14}$M$_{\odot}$, identified in each redshift subsample. The evolution of their medians in each redshift subsamples is discussed in Appendix~\ref{app:massive}.  

Despite the fact that our simulated data need a model with a curvature to describe the relation between the hot-gas fraction and the mass, we find that limiting ourself to exclusively massive systems lead to a constant and non evolving fraction in concordance with some observational results. In this limit, the fitting procedures returned almost identical values of residuals independently on the choice for the functional form and overdensity, including the power law at $R_{2500}$.

\subsection{Results on the combined sample}

%updated Dec 9th
\begin{table}
\caption{Hot-gas fraction: Best-fit parameters for Eq.~\ref{eq:glob}, Eq.~\ref{eq:stanek}, and Eq.~\ref{eq:log} with the $(1+z)$ redshift dependence parametrized as in Eq.~\ref{eq:exty} (top part) and as in Eq.~\ref{eq:extx} (bottom part). }      
\label{tab:zgas}      
\centering      
\begin{tabular}{|c |c c| c c| c c|} 
\hline
Eq.~\ref{eq:exty} & \multicolumn{2}{c|}{} & \multicolumn{2}{c|}{} & \multicolumn{2}{c|}{}\\
$f(X$)=Eq.~\ref{eq:glob} &\multicolumn{2}{c|}{$\Delta=2500$}  & \multicolumn{2}{c|}{$\Delta=500$} & \multicolumn{2}{c|}{$\Delta=200$} \\  
\hline
%valore c2: 4.5 (R2500) 0.8 (R500) e 0.61 (R200)
$A_{2,z} \ \sigma_A$ & -10.4 & 0.4 & -9.0 & 0.2 & -7.9 & 0.1\\
$B_{2,z} \  \sigma_B$ & 15.0  &0.3 & 5.3 & 0.1 & 3.0 & 0.1\\
$D_{2,z} \ \sigma_D$ & 17.3 & 0.6 & 2.5 & 0.5  & -3.4& 0.3 \\
\hline
Eq.\ref{eq:exty} & \multicolumn{2}{c|}{} & \multicolumn{2}{c|}{} & \multicolumn{2}{c|}{}\\
$f(X)$=Eq.~\ref{eq:stanek}&\multicolumn{2}{c|}{$\Delta=2500$}  & \multicolumn{2}{c|}{$\Delta=500$} & \multicolumn{2}{c|}{$\Delta=200$} \\    
\hline   
%valore c2: 4.5 (R2500)  0.77(R500) e  0.6(R200)
$A_{3,z}\ \sigma_A$ &-11.0  & 0.3 & -8.6  &    0.2  & -7.6  &  0.3   \\
$B_{3,z}\ \sigma_B$ & 8.5  & 0.6 & 5.6 &    0.4  & 4.3  &0.5 \\
$C_{3,z}\ \sigma_C$ &-6.5  & 1.5 & -3.6  & 0.4  &  -3.7 & 0.3 \\
$D_{3,z}\ \sigma_D$ &14.4  & 0.6 & 1.6 &  0.2    & -4.2 &    0.7\\
%$\chi_{3,z}$ & \multicolumn{2}{c|}{0.65}
% &\multicolumn{2}{c|}{0.56}&\multicolumn{2}{c|}{0.67}\\
% $\chi_{3,z}$ [$E(z)$] & \multicolumn{2}{c|}{[0.66]}
% &\multicolumn{2}{c|}{[0.57]}&\multicolumn{2}{c|}{[0.65]}
 %\\&0.92 &&0.93 & & 0.94&\\
\hline
Eq.~\ref{eq:exty} & \multicolumn{2}{c|}{} & \multicolumn{2}{c|}{} & \multicolumn{2}{c|}{}\\
$f(X$)=Eq.~\ref{eq:log} &\multicolumn{2}{c|}{$\Delta=2500$}  & \multicolumn{2}{c|}{$\Delta=500$} & \multicolumn{2}{c|}{$\Delta=200$} \\  
\hline
%valore c2: 4.5 (R2500) 0.8 (R500) e 0.61 (R200)
$A_{4,z} \ \sigma_A$ & -14.5 & 0.6 & -10.1  & 0.2 & -8.6 & 0.2  \\
$B_{4,z} \  \sigma_B$ & 22.2  & 1.1 &15.0 & 0.5 & 9.5 &0.6 \\
$D_{4,z} \ \sigma_D$ & 12.8 & 0.9 & 2.1 & 0.3 & -3.4 & 0.3 \\
\hline
\hline
Eq.\ref{eq:extx} & \multicolumn{2}{c|}{} & \multicolumn{2}{c|}{} & \multicolumn{2}{c|}{}\\
$f(X)$=Eq.~\ref{eq:stanek}&\multicolumn{2}{c|}{$\Delta=2500$}  & \multicolumn{2}{c|}{$\Delta=500$} & \multicolumn{2}{c|}{$\Delta=200$} \\    
\hline   
$A_{3,z}\ \sigma_A$ &-12.4 & 0.3 & -9.0  &  0.3  & -7.6  &  0.2   \\
$B_{3,z}\ \sigma_B$ & 7.7  & 1.0 & 6.1 &    0.4   & 4.0  &0.2 \\
$C_{3,z}\ \sigma_C$ &-11.7  & 0.9 & -4.1  & 0.4   &  -2.4 & 0.1 \\
$D_{3,z}\ \sigma_D$ &110.4  & 3.9 & 23.1 &  11.7    &-122.7 &    11.6\\
\hline
Eq.~~\ref{eq:extx} & \multicolumn{2}{c|}{} & \multicolumn{2}{c|}{} & \multicolumn{2}{c|}{}\\
$f(X$)=Eq.~\ref{eq:log} &\multicolumn{2}{c|}{$\Delta=2500$}  & \multicolumn{2}{c|}{$\Delta=500$} & \multicolumn{2}{c|}{$\Delta=200$} \\    
\hline 
$A_{4,z}\ \sigma_A$ &-14.8 & 0.4 & -10.2  &  0.2  & -8.3  &  0.1   \\
$B_{4,z}\ \sigma_B$ & 27.4  & 0.6 & 14.2 &    0.9   & 8.5  &0.4 \\
$D_{4,z}\ \sigma_D$ &106.0  & 2.8 & 4.7  & 26.7   &  -153.6 & 16.7 \\
\hline
\end{tabular}
\tablefoot{See notes of Table~\ref{tab:table2}. }
\end{table}

\begin{figure}
    \centering\includegraphics[width=0.49\textwidth]{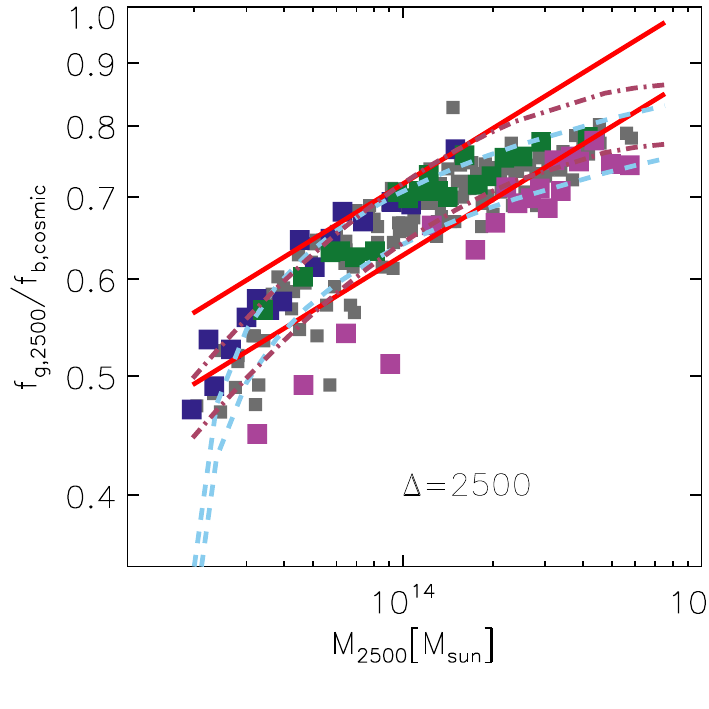}
    \caption{Medians of hot-gas fractions at $R_{2500}$. Symbols and colors have similar meaning as in Fig.~\ref{fig7}.
    The solid red, dot-dashed brown, and dashed  light blue lines represent the best-fit results of the horizontal shifting (parameters in top half of Table~\ref{tab:zgas}) for $z=0.07$ (curves with lower normalization) and $z=1.32$. }
    \label{fig:zgas}
    \end{figure}

In Table~\ref{tab:zgas} we report the means and standard deviations of the 1000 iteration best-fit parameters of the linear, parabolic and logarithmic descriptions once the redshift dependence is included in both Eq.~\ref{eq:exty} and Eq.~\ref{eq:extx} as $(1+z)$ and the analysis is applied simultaneously to all redshift bins (we omit the linear fit in the second part of the table because the two shifts in this case coincide). 
Both the parabola and the logarithmic formulae are well suited to describe the mass dependence of the hot-gas fraction at $R_{2500}$ as they were for the total baryon fraction as seen in Fig.~\ref{fig:zgas}.

\begin{figure}
\includegraphics[width=0.49\textwidth]{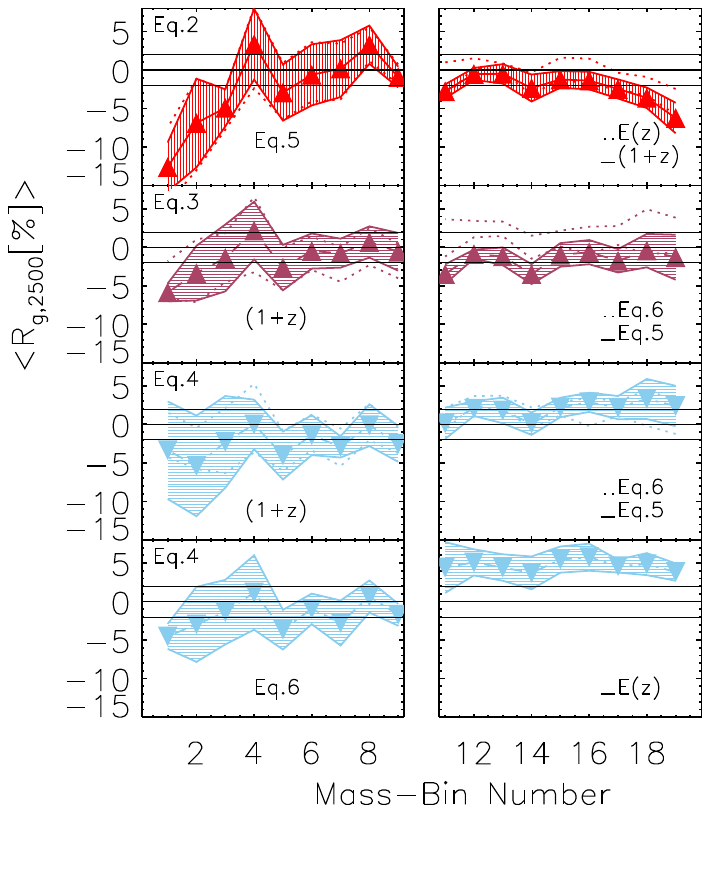}
\caption{Average and 1$\sigma$ deviation of the residuals of the gas fraction at $R_{2500}$ at fixed mass-bin number. Colors and lines as in Fig.~\ref{fig:resb}. }
\label{fig:resgas}
\end{figure}

In Fig.~\ref{fig:resgas}, we report the residuals of the hot-gas fraction for all the cases shown before for the baryon fraction. The biggest difference is that all models deviate more from zero at the lowest mass bins. In addition, the advantage of using $E(z)$ in Eq.~\ref{eq:extx} for the logarithmic function is no more present. Indeed, in the bottom panel, neither the residuals of the first mass bins are zero nor the dispersion is smaller than the third panel.  The explanation is most likely due to the failure of the theoretical premises for which the gas fraction follows a self-similar trend. In case of non-radiative simulations, the hot-gas fraction does not experience any drop at low-masses and is faithfully reproduced by a power-law form for which a translation along the $X$ axis coincides with a translation along the $Y$ axis. However, in the presence of cooling/star-formation and heating from feedback sources, the hot-gas fraction-mass relation acquires a curvature that for the highest overdensity depends more on mass than on redshift.  A clear visualization of this is the right panel of Fig.~\ref{fig:barpresentation} where the four redshift sub-samples behave almost as a unique population that changes its trends with $M_{200}$ between $2$ and $3 \times 10^{14} M_{\odot}$, independent of the redshift. The evolution of the gas fraction of individual objects in this plane depends indeed on their high-z mass. If they are massive systems ($M_{200}> 2\times 10^{14}$M$_{\odot}$) already at $z\ge 1$ during their evolution they will just shift horizontally. Vice versa, if at high-redshift they had a smaller mass, their mass substantially evolves, either due to mergers or diffuse accretion. Their associated radius will be larger and most of the hot gas that was previously expelled by feedback will be recounted within the cluster boundaries. Therefore, for these objects not only the total mass grows, but also their hot-gas fraction. The different evolution of the gas fraction for small and massive high-redshift clusters impacts more at the highest overdensity, and explains why a pure mass-shift is less suitable to account for the evolution of the scaling relation at $R_{2500}$. If our sample was to extend to even smaller masses or if the simulations would have stronger AGN feedback, the effect will be stronger, as presented below.

For the residuals at the lowest overdensity, we notice that in case of the hot-gas fraction, the linear fitting is poorly performing in the first 2 bins even at $R_{500}$ showing deviation around -5 percent. At $R_{200}$, the deviations in the first bins are still present but of smaller amplitude.

%--------------------------------------------
%--------------------------------------------
\section{Comparison with other simulations}
\label{sec:compasim}
%--------------------------------------------
%--------------------------------------------

\begin{figure*}
%    \centering
   \includegraphics[width=0.95\textwidth]{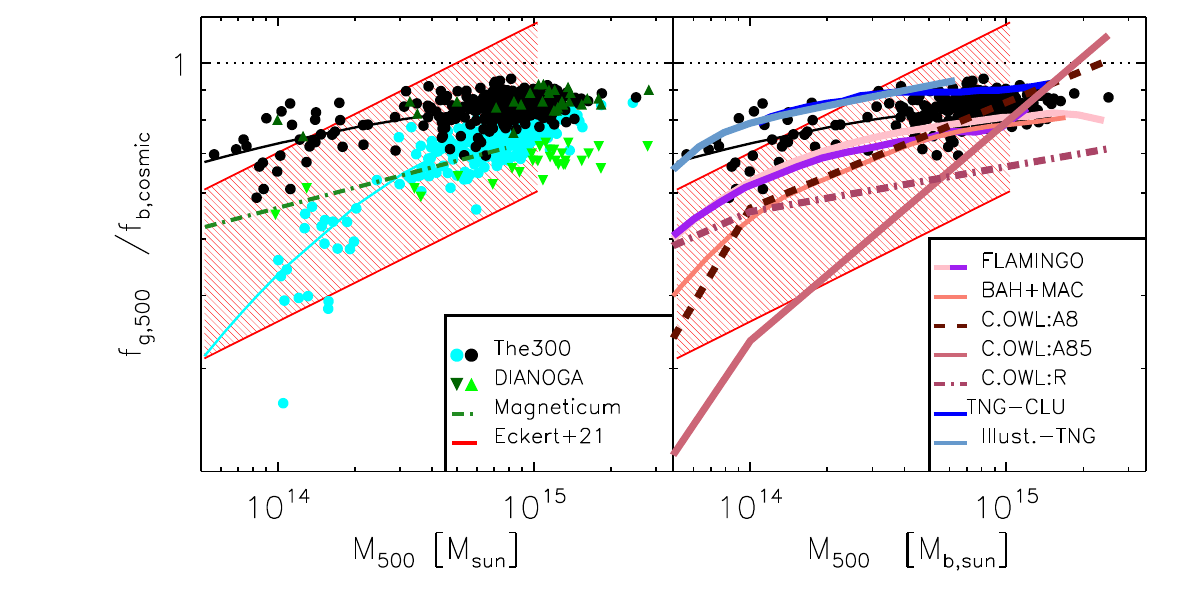}
    \caption{Comparison between different simulation sets for the gas fraction relative to the cosmic baryon fraction within $R_{500}$ versus the total cluster mass, $M_{500}$, at $z=0$. In both panels, the red band is the observational reference from \cite{eckert.etal.2021} (as Fig.~\ref{fig:fig2}) and the gray empty circles shown The300 simulated clusters used in this work. In the left panel, the relative gas fraction from: The300 simulated with {\tt GIZMO-SIMBA} in cyan; the AGN and CSF models of the {\tt DIANOGA} sets in dark and bright green triangles. The median values for {\tt TNG-Cluster} are in blue; those for the largest and the highest resolution box of the {\tt FLAMINGO} simulations in pink and the purple; the combination of the {\tt BAHAMAS} and {\tt MACSIS} samples is in orange. In the right panels, we report some best-fit forms: those for {\tt cosmo-OWL} AGN8, AGN85, and reference model are in dark brown, brown, and light brown. The model for {\tt Illustris TNG} is in light blue and that for the {\tt MAGNETICUM} run in black (appropriate references in the text).}
    \label{fig:sim}
    \end{figure*}
    
%--------------------------------------------

In \cite{cui.etal.2022} the gas fraction measured in the simulated clusters used in this work was compared with other simulation sets, such as {\tt FABLE} \citep{henden.etal.2020}, {\tt C-EAGLE} \citep{barnes.etal.2017ceagle}, and the {\tt GIZMO-SIMBA} version of \tre. Their figure~1 (left panel) shows the excellent agreement among all simulated results, with the exception of an overall lower gas fraction for the {\tt GIZMO-SIMBA} set, that becomes more evident at the lowest masses reaching a relative gas fraction of $0.4$ at $10^{14}$M$_{\odot}$. We present here some further comparisons with other simulation datasets. We discuss in the next section the possible origin for the differences with the {\tt GIZMO-SIMBA} clusters, which are generated from the same initial conditions as the simulations analyzed here, and with {\tt MAGNETICUM}, which have been realized with a {\tt GADGET-3} code version that implemented the same {\tt GADGET-X} hydrodynamics \citep{beck.etal.2016} and that assumes the same prescriptions for star formation and stellar evolution.

\vspace{5pt}

In Fig.~\ref{fig:sim}, we report a comparison between both versions of The300 clusters and other sets of simulations. In the left panel, we include the $z=0.07$ sample (black points), together with the same clusters simulated with the {\tt GIZMO-SIMBA} code (cyan points), the AGN and CSF sets from {\tt DIANOGA} \citep{planelles.etal.2017} (upward and downward triangles), and the best-fitting relation of the groups and clusters from Box2b of the {\tt MAGNETICUM} suite \citep{angelinelli.etal.2023}. In the right panel, we compare our results with the average baryon fraction for {\tt FLAMINGO} \citep{schaye.etal.2023} and {\tt TNG-Cluster} \citep{nelson.etal.2023}, which coincides with that of {\tt Millenium-TNG} \citep{pakmor.etal.2023} not shown. We also consider the best-fit relations of $(i)$ three simulation sets from {\tt cosmo-OWL} \citep{lebrun.etal.2017}, $(ii)$ a combination of {\tt BAHAMAS} and {\tt MACSIS} groups and clusters \citep{farahi.etal.2018}, $(iii)$ galaxy, groups and clusters from {\tt Illustris-TNG} \citep{pop.etal.2022}. In the case of {\tt cosmo-OWL} and {\tt Illustris-TNG} we report the best-fitting broken-power-law and smoothly-broken-power-law relations \citep[see][respectively]{lebrun.etal.2017,pop.etal.2022}. To ease the comparison with Fig.~\ref{fig:fig2}, we also show the red band from \cite{eckert.etal.2021} which encompasses several observational results.

In this plot, we rescale the masses by the different values for the Hubble constant, and the gas fractions by the different values of the cosmic baryon fraction assumed in each simulation set. These corrections are, however, minimal; all simulations, with the exception of {\tt DIANOGA}\footnote{The {\tt DIANOGA} clusters assume a reduced Hubble parameter $h=0.72$ and a cosmic baryon fraction of 0.167.}, assume a Planck cosmology with differences on the relevant parameters which are only in the third digit. 

Not surprisingly, the gas fraction of the {\tt DIANOGA-AGN} (dark green upward triangles) is very similar to that of the \tre\ since they are performed with the same code and a similar implementation of AGN feedback. 
Assuming the same physics for radiative cooling, star formation, and stellar evolution, but excluding the AGN feedback substantially reduces the gas fraction, as seen by comparing the {\tt DIANOGA-AGN} with the {\tt DIANOGA-CSF} set (green downward triangles).
%Excluding the AGN feedback substantially reduces the gas fraction {\bf as seen by comparing this set with the {\tt DIANOGA-CSF} (green upward triangles), which include the same physics for radiative cooling, star formation, and stellar evolution. 
A similar conclusion can be drawn by comparing the only-radiative sample (R)  of {\tt cosmo-OWL} (dark-red dot-dashed line) to the other two sets of the same simulation suite that include AGN feedback. Also in this case, when the radiative cooling is not balanced by the AGN feedback within massive halos, an exceedingly large fraction of the gas is converted into stars, thus not contributing to the hot-gas fraction. The {\tt cosmo-OWL} sets with AGN (with AGN85 feedback being more powerful than for AGN8) have a reduced gas fraction with respect to the \tre\ sample at almost all mass scales, with a difference that {increases as the mass reduces and expands} for more efficient AGN feedback. Indeed, gas fractions for the {\tt cosmo-OWL}, the {\tt BAHAMAS+MACSIS}, and {\tt MAGNETICUM} are around 0.6 at $10^{14}$M$_{\odot}$, in agreement with observational data. Nevertheless, the broken-power-law trend applied to the {\tt cosmo-OWL} data with AGN does not capture the observed flattening of the gas fractions at large cluster masses that characterizes the observational data (see Fig.~\ref{fig:fig2}). On the other hand,  {\tt BAHAMAS}+{\tt MACSIS}, {\tt MAGNETICUM}, and {\tt FLAMINGO} seem to reach a plateau in the high-massive regime which converges at $f_{g,500} \simeq  0.8$, and thus approximately ten percent below our value.
%, it is slightly below the observational results for clusters of mass $M_{500}=10^{15}$M$_{\odot}$  which is closer to $0.9$ (see, for example, \citealt{lyskova.etal.2023} for a comparison with CHEX-MATE and \citealt{biffi.etal.2025} for a comparison with X-COP).
%Nevertheless, it has to be noted that all simulations fall in any case within the red band plotted in the figure. 
Finally, the smoothly broken power-law description of {\tt Illustris-TNG}, quite similar to that of {\tt TNG-Cluster},  shows the same trend of our data -- and to those of the {\tt FLAMINGO} suite -- with a slightly larger normalization, which locates their relation on the upper side of the observational results. %Finally, the  {\tt FLAMINGO} simulations show a curvature similar to our data and to {\tt TNG-CLUSTER}, but with a lower normalization, independently of the sampled volume and of the numerical resolution.

\section{Discussion}
\label{sec:discussion}

In this Section, we analyze in more depth the comparisons shown in the left panel of Fig.~\ref{fig:sim} and begin with the difference between the two versions of The300: {\tt GADGET-X} and {\tt GIZMO-SIMBA}. We recall that {\tt GIZMO-SIMBA}  is characterized by a high jet velocity of the AGN feedback that at high redshift expels the gas, pushing it away from the boundary of the halos (see \citealt{li.etal.2023}, which presents the evolution of all the gas phases - cold, warm, hot - as well as the stellar component from $z=0$ to $z=4$).  One consequence of this behavior of the AGN feedback is that the brightest cluster galaxies (BCGs) are efficiently quenched. We note that the {\tt GIZMO-SIMBA} BCGs are, nevertheless, already significantly massive at $z=2$ as a result of the efficient star formation model implemented in that code. Thus, despite the AGN feedback prevents star formation at low redshift, the {\tt GIZMO-SIMBA} BCGs are typically more massive and with more concentrated stellar cores even at low redshifts with respect to the {\tt GADGET-X} BCGs \citep{meneghetti.etal.2023,srivastava.etal.2024}. The gas, efficiently expelled in {\tt GIZMO-SIMBA}, becomes completely unbound, so that it is prevented from being reaccreted onto the clusters at a later time. This gas will eventually enrich the diffuse gas that permeates the large-scale cosmic web (see, for example, Figure 2 of \citealt{li.etal.2025} showing that {\tt GIZMO-SIMBA} has higher gas density profiles at $R>2 R_{200}$ once the subhalos contribution is removed.)

The comparison with the results from {\tt MAGNETICUM} is more subtle. In fact, as mentioned above, {\tt MAGNETICUM} has some common features with {\tt GADGET-X}, while having at the same time also key differences on the conduction model, on the details of the BH seeding and of AGN feedback (see, \citealt{sembolini.etal.2016}, for the description of both codes). Such variations in the model make it difficult to understand the main reason for the differences in the resulting gas fraction. 
{\tt GADGET-X} uses artificial thermal diffusion \citep{beck.etal.2016} to better capture gas-dynamical instabilities, while {\tt MAGNETICUM} adopts a physical thermal conduction set at $1/20^{\rm th}$ of the Spitzer conduction. In massive systems, these two prescriptions produce similar effects of homogenizing the ICM by suppressing the presence of low-entropy clumps. Their greatest impact is on the thermal complexity of the ICM \citep{dolag.etal.2004, rasia.etal.2014}, but they are not expected to influence the gas fraction of massive systems. 

For the BH/AGN sector, the switching between the quasar and radio modes in {\tt GADGET-X} is modeled following \cite{Steinborn2015} which has been shown to produce a realistic quenching of galaxies at high redshift \citep{chittenden.etal.2025} and a reduced gas fraction at $z>2$ \citep{dolag.etal.2025}. Such features are not recovered in {\tt MAGNETICUM} that follows the prescription proposed by \cite{fabjan.etal.2010} for the transition from quasar to radio  mode\footnote{The paper \cite{dolag.etal.2025} presents the {\tt MAGNETICUM} Pathfinder set of simulations and includes the comparison of the gas fraction measured in one small box simulated up to $z\sim2$ both with the standard {\tt MAGNETICUM} prescription by  \cite{fabjan.etal.2010} and with the one proposed by \cite{Steinborn2015}.}. 

On the other hand, in {\tt GADGET-X} the BHs are repositioned at each time step at the location of the most-bound neighbor particle within their softening lengths, while in {\tt MAGNETICUM} they are assigned a large dynamical mass to artificially boost the otherwise unresolved effect of dynamical friction. Pinning on the local potential minimum usually leads to a more bursty AGN feedback at high redshift, which might provide more ICM heating and avoid later gas cooling. 
However, in The300 simulations the BH softening length is relatively larger than in {\tt MAGNETICUM}, thus causing the BHs to wander, especially at relatively low redshift, $z\lesssim 1$,
when the growth of the BCG \citep{ragone-figueroa.etal.2018,contreras-santos.etal.2022} and their central BH \citep{bassini.etal.2019} is influenced by major mergers with other massive galaxies.
%when the most massive systems are subject to an increase number of major mergers directly affecting the growth of the BCG \citep{ragone-figueroa.etal.2018,contreras-santos.etal.2022} and the central BH \citep{bassini.etal.2019}. 
Under these dynamical conditions, the central BHs, locked in the potential minimum, can be displaced from the BCG's center of mass or from the galaxy itself, in case of a merging that brings a galaxy with a lower potential minimum closer to the BCG. As an alternative, the approach based on boosting BH dynamical masses, as implemented in {\tt MAGNETICUM}, mimics the effect of unresolved dynamical friction \citep{bassini.etal.2020}, although it misrepresents the BH dynamics \citep{chen.etal.2022,damiano.etal.2024}. Being the {\tt GADGET-X} central baryon cycle not regulated anymore by the AGN feedback when the BH is not in the BCG's center, radiative cooling takes over, thereby making the BCG still actively producing stars  even at low redshift \citep{li.etal.2020,cui.etal.2022}, and inducing a flow of the low-entropy gas towards the center. This is also witnessed in the change in gas density profiles that from $z=1$ to $z=0$ reach a denser peak \citep{li.etal.2023}. 

Going beyond the simple global measure of the gas fraction, the hot gas density profiles predicted by the {\tt GADGET-X} version of The300 simulations are generally in better agreement with the observations of both local ($z\sim0$, \citealt{mcdonald.etal.2017,ghirardini.etal.2019}) and high-redshift ($1.2>z>1.7$, \citealt{mcdonald.etal.2017}) clusters \citep{li.etal.2023} than the {\tt GIZMO-SIMBA} clusters. 
%Recent comparisons, extended to the CHEX-MATE sample, also favor {\tt GADGET-X} with respect to the ICM properties, such as the X-ray emission-measure profile \citep{bartalucci.etal.2023}, temperature profiles \citep{rossetti.etal.2024}, and  entropy profiles \citep{riva.etal.2024}.

\vspace{5pt}

Before concluding, we would like to stress that the differences among the various sets of simulations do not impact our main result, which concerns the ability of our model to adequately fit the largest variety of trends of the total baryon and hot gas fractions against the total cluster mass. In fact, all simulations plotted in Fig.~\ref{fig:sim} show
trends similar to those described in our analysis: a flattening of the baryon and hot gas fractions at high masses and a concave curve from the cluster towards the group scale. 
%, are features common to several other simulations. 
The curvature and fraction reduction for groups is more pronounced in the case of strong gas depletion caused by outflows driven by AGN feedback, which affects smaller mass systems.  However, even for the clusters selected from {\tt GIZMO-SIMBA} version, i.e., carried out with the code that presents one of the most efficient feedback models, our quadratic formula adequately describes the gas fraction. This is shown by the solid cyan line in the left panel of Fig. \ref{fig:sim}, which represents the best-fitting parabolic form (with $A=19.2$, $B=27.3$, and $C=-17.3$). Therefore, although the parameters of the fitting functions might differ, the proposed functional forms seem suitable to describe the predictions of virtually all state-of-the-art cosmological hydrodynamical simulations that provide a large sample of simulated clusters and groups.

\section{Conclusion}
\label{sec:concl}
%--------------------------------------------
%--------------------------------------------

We present the evolution of the baryon and hot gas fractions in massive galaxy clusters from $z\sim 0$ to $z\sim 1.3$, as predicted by The \tre \ suite of cosmological hydrodynamical simulations \citep{cui.etal.2018} carried out with the {\tt GADGET-X} code, and presented in Sect.~\ref{sec:sim}. 
The purpose of our analysis is determine a highly flexible analytic description of the relation between these fractions and the total cluster mass, which is accurate at all redshifts and at the three considered overdensities ($\Delta=2500, 500$ and $200$). To this purpose, we compare the usual power-law representation (Eq.~\ref{eq:glob}) with a parabolic (Eq.~\ref{eq:stanek}) and logarithmic (Eq.~\ref{eq:log}) functions in a logarithmic plane, similar to those adopted in \cite{stanek.etal.2010} and \cite{vikhlinin.etal.2009}, respectively. Furthermore, we investigate three alternative formulations to describe the evolution of such fractions (Eqs.~\ref{eq:exty}-\ref{eq:Zpla}), by considering two alternative expressions for the redshift dependence, $(1+z)$ and $E(z)$. The performance of small variations of the parametric formulae is also quantitatively assessed (see Appendix~\ref{app:extra}). 

The baryon/hot-gas fraction-mass relations from the simulations analyzed here favorably compare with observational results on individual clusters (see Fig.~\ref{fig:fig2}), although the results from simulations have a smaller scatter. This difference is possibly due to uncertainties in the observational estimates and/or a limited capability of the star formation and feedback processes included in the simulations to describe the diversity of the cosmic baryon cycle in different clusters (see Sect.\ref{sec:obse}).

Our main results can be summarized as follows.

   \begin{enumerate}
      \item Baryons, and in particular the hot gas, are more depleted in the core regions, within $R_{2500}$, where their fractions have also a larger scatter (Figs.~\ref{fig:barpresentation}, \ref{fig6}, \ref{fig7}, and \ref{fig:zgas}). This is mostly caused by AGN feedback that has more effect on the smaller masses and {\bf is} more powerful at higher redshifts. The mass-dependence of the baryon fraction, and even more of the hot-gas fraction, has three main characteristics: a flat behavior at the highest masses, followed by a curvature at intermediate masses, and a drop when reaching the scale of galaxy groups. 
      \item When forcing a power-law description on the relation between the baryon fraction and the total mass, both slopes and normalizations  decrease with time at $R_{2500}$, while they remain constant at lower overdensities (Sect~\ref{sec:powelaw}). The slope $b_2$ of the power-law relation (Eq.~\ref{eq:glob}) is higher in the inner regions. For the cluster mass range sampled by our simulations, $R_{2500}$ varies from $\sim 280$ kpc in groups to $~875$ kpc in clusters at $z=0$. Thus, while AGN feedback has a small impact on the total baryon content within the most massive clusters, it is quite effective in removing gas from the core regions of the smallest objects. This is even more evident at high redshifts when the 10 percent smallest objects that we consider have $R_{500}\lesssim 150$ kpc. In any case, our results show the ineffectiveness of a power law function to describe the relation between the baryon/hot-gas fraction and mass (see Fig.~\ref{fig6}).
      \item The curved trend of the baryon/hot gas-mass relations is better described by a parabola or a logarithm in a logarithmic plane (see Figs.~\ref{fig4}--\ref{fig:resgas}). By comparing the residuals of these three different functional forms, we find that largest deviations are always associated to the power-law, as a consequence of a misrepresentation either at the low-mass or at the high-mass regimes (see Fig.~\ref{fig:resb} and Fig.~\ref{fig:resgas}). On the other hand, in the mass range considered here both the parabola and the logarithmic expressions can efficiently and accurately describe at the same time the drop of the baryon fraction in groups, and the flattening at highest mass, although the logarithmic functional form still slowly grows for the most massive objects, thus not exactly capturing the asymptotic plateau. The logarithm form has the advantage to have only two free parameters, as the linear fit, once the vertical asymptote is fixed, which is the preferred choice (see Appendix~\ref{app:extra}). Both the parabola and the logarithm expressions are accurate at the highest overdensity, and for the hot gas fraction. In summary, at fixed redshift, we recommend to use one of the following expressions: \\  
      
      $\log(f/f_{\rm cosmic})=a+b\log(M/M_0)+c(\log(M/M_0))^2$ or \\
      $\log(f/f_{\rm cosmic})=a+b\log(\log(M/M_0+1.2))$ \\
      where $M_0=3\times 10^{14}$M$_{\odot}$ with the parameters taken from Table~\ref{tab:eq2} and Table~\ref{tab:eq3}, respectively, for the baryon fraction and Table~\ref{tab:eq2gas} and Table~\ref{tab:eq3gas} for the hot-gas fraction (the parameter of the quadratic form for the $z=0.07$ {\tt GIZMO-SIMBA} set are listed at the end of in Sect.~\ref{sec:discussion}).
      \item As for the evolution, we find that it can be modelled by using either $(1+z)$ or $E(z)$ to express the redshift dependence, with the former being slightly preferred by the goodness of the fit. 
      When analyzing the combined sample, the evolution of the baryon fraction at $R_{2500}$ is better parameterized by the parabola adding the evolution factor to either the normalization or the mass (see discussion in Sect.~\ref{sec:res_mz} and Fig.~\ref{fig:resb}). The logarithm, instead, performs better, reducing the residuals of the first bins to zero, for the baryon fraction when the evolution consider the $E(z)$ factor to be added to the total mass. For the gas fraction, instead, the evolution is well capture by adding an extra term depending on $(1+z)$ to the normalization for either the parabolic and logarithmic forms. The logarithm, however, performs less suitably due to a marked redshift dependence of the curvature which cannot be well represented when we fit the combined sample.  From our analysis, we therefore suggest to use the following expression:\\
      
      $\log(f/f_{\rm cosmic})=a+b\log(M/M_0)+c(\log(M/M_0))^2+d \log((1+z)/(1+z_0))$, \\
      where $M_0=3\times 10^{14}$M$_{\odot}$, $z_0=0.53$, and the parameters are taken from second blocks of  Table~\ref{tab:table2} and  Table~\ref{tab:zgas}, respectively, for the baryon and hot-gas fractions. At $R_{500}$ and $R_{200}$ this expression and the logarithmic one are equally suitable.
      \item The bending of the baryon/hot-gas fractions at decreasing cluster masses, that we found in our analysis, is common to all state-of-the-art cosmological simulations, and it is even more pronounced when the smallest mass systems (groups and below) are included.
      
   \end{enumerate}

Regarding the last point, it is worth remarking that simulations that focused on the group scales, such as EAGLE \citep{davies.etal.2020,Oppenheimer.etal.21}, HYNEAS \citep{cui.etal.24}, or X-FABLE \citep{bigwood.etal.2025} find a low hot-gas fraction for systems with $M_{500}<10^{14}M_{\odot}$. Therefore, even though the exact values of the parameters of the fitting formulae presented in this paper might depend on the specific simulation analyzed, we expect that the described trends hold for systems at $10^{14}$M$_{\odot}$ or slightly below. For even smaller systems, with mass $(1-5)\ 10^{13}$M$_{\odot}$, different simulations predict significantly different results for the baryon/hot-gas fractions. This can be seen in Fig.~\ref{fig:sim}, especially with respect to the best-fitting relations by \citealt{farahi.etal.2018} for the {\tt BAHAMAS-MACSIS} sample and by \citealt{angelinelli.etal.2023} for {\tt MAGNETICUM}. The latter work has recently being favorably compared with eROSITA data of groups with $M_{500}$ of about and below $10^{14}$M$_{\odot}$ \citep{popesso.etal.2024arxiv}.  Values as small as $\approx  5$ percent in this mass range are in line with some of the latest observational results \citep[ see also,][]{andreon.etal.2024,marini.etal.2024, popesso.etal.2024}. In fact, these analyses pointed out that the values presented in the literature for the hot gas fraction of the group-scale objects could be biased high by selection effect, being the gas-dense objects more easily detected in the X-rays. 

For these low-mass systems, the observational measurements of the total mass is even more challenging. Indeed, being gas-poor their X-ray and SZ signal is weak and does not extend to large radii; their richness can be easily affected by the presence of projection interlopers; their mass can hardly be measured by gravitational lensing. On the other hand, cosmological surveys both in the X-rays (e.g., e-ROSITA; \citealt{merloni.etal.2024,bulbul.etal.2024}) and in the optical/near-IR (e.g., Euclid; \citealt{mellier.etal.2024}) will provide a huge number of groups, thus enabling statistically robust analyses using stacking techniques such as that performed in \cite{popesso.etal.2024}. However, the development of new techniques would be needed to achieve robust results in high-redshift objects, because stacking of Euclid high-$z$ clusters and groups in eROSITA are not expected to effectively constrain group-scale gas fractions due to the point-spread-function limitations which could lead to an large miscentering.
Given that the sensitivity of the predicted hot gas and baryon fractions depends on the details of the galaxy formation models implemented in different simulations, forthcoming observational results at the group scale will have a crucial role to shed light on our understanding of the cosmic baryon cycle.

\begin{acknowledgements}
We thank the Referee for providing useful comments.
We thank Adam Mantz for sharing his results on the gas and total mass from \cite{mantz.etal.2016}; Roi Kugel and Joop Shaye for sharing the FLAMINGO gas fraction used in Fig.~\ref{fig:sim}; R\"udiger Pakmor and Dylan Nelson for sharing the Millenium-TNG and Cluster-TNG gas fraction used in the same figure.
ER, VB, and CA acknowledge support from the Chandra Theory Program (TM4-25006X)
awarded from the Chandra X-ray Center which is operated by the Smithsonian
Astrophysical Observatory for and on behalf of NASA under contract NAS8-
03060.
ER and VB acknowledge partial support from the INAF Grant 2023 ``Origins of the ICM metallicity in galaxy clusters". 
This paper is supported by: the Italian Research Center on High Performance Computing Big Data and Quantum Computing (ICSC), project funded by European Union - NextGenerationEU - and National Recovery and Resilience Plan (NRRP) - Mission 4 Component 2, within the activities of Spoke 3, Astrophysics and Cosmos Observations; the National Recovery and Resilience Plan (NRRP), Mission 4,
Component 2, Investment 1.1, Call for tender No. 1409 published on
14.9.2022 by the Italian Ministry of University and Research (MUR),
funded by the European Union – NextGenerationEU– Project Title
"Space-based cosmology with Euclid: the role of High-Performance
Computing" – CUP J53D23019100001 - Grant Assignment Decree No. 962
adopted on 30/06/2023 by the Italian Ministry of Ministry of
University and Research (MUR); by the INAF project "CONNECTIONS" (COllaboratioN oN codE development for future Cosmological
simulaTIONSa INAF Grant within the "Astrofisica Fondamentale" funding scheme; by the INFN Indark Grant.
WC is supported by the Atracci\'{o}n de Talento Contract no. 2020-T1/TIC-19882 granted by the Comunidad de Madrid in Spain and the science research grants from the China Manned Space Project. He also thanks the Ministerio de Ciencia e Innovaci\'on (Spain) for financial support under Project grant PID2021-122603NB-C21 and ERC: HORIZON-TMA-MSCA-SE for supporting the LACEGAL-III project with grant number 101086388.
MDP acknowledges financial support from PRIN-MIUR grant 20228B938N "Mass and selection biases of galaxy clusters: a multi-probe approach" funded by the European Union Next generation EU, Mission 4 Component 1  CUP C53D2300092 0006.
KD acknowledges support by the Deutsche Forschungsgemeinschaft (DFG, German Research Foundation) under Germany’s Excellence Strategy - EXC-2094 - 390783311 and through the COMPLEX project from the European Research Council (ERC) under the European Union’s Horizon 2020 research and innovation program grant agreement ERC-2019-AdG 882679.
SE acknowledges the financial contribution from the contracts
Prin-MUR 2022 supported by Next Generation EU (M4.C2.1.1, n.20227RNLY3 {\it The concordance cosmological model: stress-tests with galaxy clusters}), and from the European Union’s Horizon 2020 Programme under the AHEAD2020 project (grant agreement n. 871158). 
MG acknowledges support from the ERC Consolidator Grant \textit{BlackHoleWeather} (101086804). 
This work has been made possible by the The Three Hundred collaboration\footnote{https://www.the300-project.org} and 
The authors acknowledge The Red Espa\~nola de Supercomputaci\'on for granting computing time for running the hydrodynamic simulations of The300 galaxy
cluster project in the Marenostrum supercomputer at the Barcelona
Supercomputing Center.

\end{acknowledgements}

% WARNING
%-------------------------------------------------------------------
% Please note that we have included the references to the file aa.dem in
% order to compile it, but we ask you to:
%
% - use BibTeX with the regular commands:
%   \bibliographystyle{aa} % style aa.bst
%   \bibliography{Yourfile} % your references Yourfile.bib
%
% - join the .bib files when you upload your source files
%-------------------------------------------------------------------

\bibliographystyle{aa}

\begin{appendix}

\section{Baryon fraction: Analysis and Tables}
\label{app:bar}

In the following, we present the detailed results of our study on the modeling of the relation between the total mass and the baryon fraction. In Figs.~\ref{fig3},~\ref{fig4}, and \ref{fig5} as well as in Tables~\ref{tab:eq1}, \ref{tab:eq2}, \ref{tab:eq3}, we show and report the means and standard deviations of the 1000 iterations for the best-fitting parameters of Eqs.~\ref{eq:glob}, \ref{eq:stanek}, and \ref{eq:log}.

%--------------------------------------------
\subsection{Linear form}
\label{sec:powelaw}
%--------------------------------------------
%\ER{DONE with this}
\begin{figure}
    \centering
\includegraphics[width=0.49\textwidth]{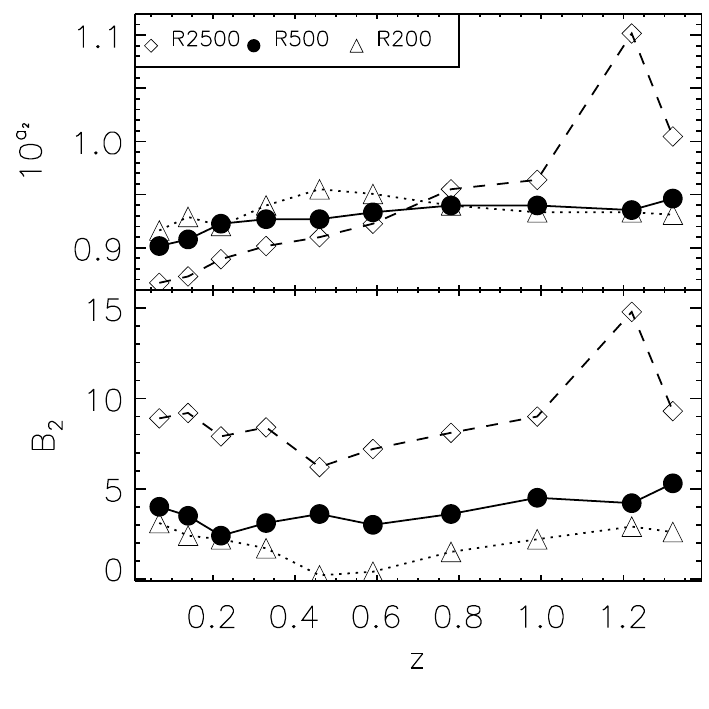}
    \caption{Best-fit parameters of Eq.~\ref{eq:glob} (linear relation in the logarithmic plane) for the redshift sub-samples. Dashed, solid, and dotted lines and empty diamonds, solid circles, and empty triangles respectively refer to $\Delta=2500, 500$, and $200$. Error bars on the parameters are smaller than the symbols. 
    }
    \label{fig3}
\end{figure}

 The linear fit in a logarithmic plane (Eq. \ref{eq:glob}) results from the baryon mass being expressed as a power law of the total mass. For this, we expect the slope to be positive ($b>0$).

The evolution of the normalization (Fig.~\ref{fig3}) is present when the quantities are measured at $R_{2500}$ (dashed line; see also Table~\ref{tab:eq1}). The variation of the $a$ parameter seems in contrast with the previous claim about an almost absent evolution of the median baryon fraction measured for the entire sample from $z\sim 1$ to $z\sim 0$. However, we should remember that the normalization, $10^a$, shows the relative fraction values of the best-fit relation at $M_0=3\times 10^{14}$\msun, a mass not covered by the higher redshift samples and we should note that exactly these sub-samples are characterized by a steeper relation (see below). The combination of these two facts leads to a higher values of $a$ at higher redshift, still without implying that the relative baryon fraction actually decreases towards $z=0$ (as evident from Fig.~\ref{fig:barpresentation}).  
 Using the {\sc idl} routine {\sc robust\_linfit} we find that the evolution of the normalization of the baryon fraction at $R_{2500}$ can be described as
 \begin{equation}
     10^{a_2}=0.75 + 0.11\, (1+z),
     \label{eq:evnorm}
 \end{equation}
 with $1\sigma$ uncertainties lower then $0.005$ on both intercept and slope. The relation between $10^a$ and $E(z)$ has the same intercept and a slightly steeper slope (equal to $0.12$). The errors associated to the parameters are twice as large in case of $E(z)$ but are still $\lesssim 0.01$. 
 
The slope $b_2$ of the relation between the logarithmic of the relative fraction and the mass is higher when the measurements are done closer to the core (within $R_{2500}$), as shown by the dashed line in the middle panel of Fig.~\ref{fig3}. This is consistent with Fig.~\ref{fig:barpresentation} and with the idea that most of the depletion in the central regions is caused by the effect of AGN feedback, which is relatively more efficient in removing gas from the core regions of smaller objects. This effect should be more evident at high redshift for three reasons. First because of the construction of our sample built to follow the evolution of massive clusters  at $z>1$ we reach smaller objects; secondly, even at parity of potential well, the AGN feedback is expected to be relatively stronger in distant clusters, and lastly the overall cluster volume measured with respect to the critical density is smaller, so that the gas expelled from the cluster center might be located well beyond $R_{2500}$ at $z>1$. That said, in our sample, the relation between the slope, $b$, and the redshift is extremely shallow even in the cluster cores and it is consistent with zero at 2$\sigma$: $b_2=(0.05 \pm 0.02)+(0.02 \pm 0.01)(1+z)$. 
 
The normalizations and slopes derived at $R_{500}$ and, in particular, at $R_{200}$ are almost redshift independent. 
 Looking at the right panel of Fig.~\ref{fig:barpresentation}, we might understand this behavior. The majority of our objects has a mass $M_{200}$ well above $10^{14}$ \msun\, even at the highest redshift. On average, thus, even at $z=1.32$ our simulated clusters have a $R_{200}$ as large as $\sim 750$ kpc. 
 The AGN feedback might not be able to efficiently displace the gas beyond such radii, or the previously expelled gas could have been already re-accreted in the external regions of these massive systems.

%--------------------------------------------
\subsection{Parabolic form}
\label{sec:parabola}
%--------------------------------------------
\cite{stanek.etal.2010} introduced the parabolic form (Eq.~\ref{eq:stanek}) to describe the relation between the gas fraction and the total mass of clusters selected in the mass range of $0.5<M_{500}/ [10^{14}$ \hinv \msun$]<10$ in a simulated cosmological volume where a strong preheating was included to mimic the effect of AGN feedback at high redshift. We employ this description because the baryon fraction as measured in large simulated samples grows rapidly with the cluster mass up to $M\approx (2$--$3)\times 10^{14}$ \msun\  \citep[see also][]{farahi.etal.2018} and it becomes almost constant at higher masses, as shown in Fig.~\ref{fig:barpresentation}. We therefore expect this trend to be consistent with the shape of a parabola with a small concave curvature and truncated at a location near its vertex. If $c_3=0$ in Eq.~\ref{eq:stanek}, the expression reduces again to a linear scaling. Therefore, we note that $b_3$ can not be negative when $c_3=0$. 

\begin{figure}
    \centering
   \includegraphics[width=0.49\textwidth]{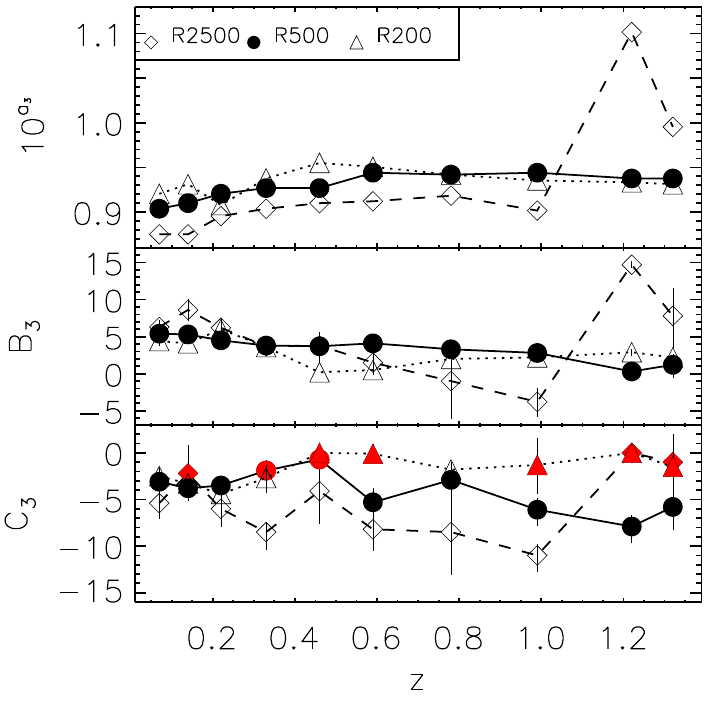}
    \caption{Best-fit parameters of Eq.~\ref{eq:stanek} (parabola in the logarithmic plane) for the redshift subsamples. Red symbols indicate $c_3$ equal to zero, or consistent within $1\sigma$. 
    The meaning of symbols and lines is similar to Fig.~\ref{fig3}.}
    \label{fig4}
\end{figure}

\begin{figure}
    \centering
   \includegraphics[width=0.49\textwidth]{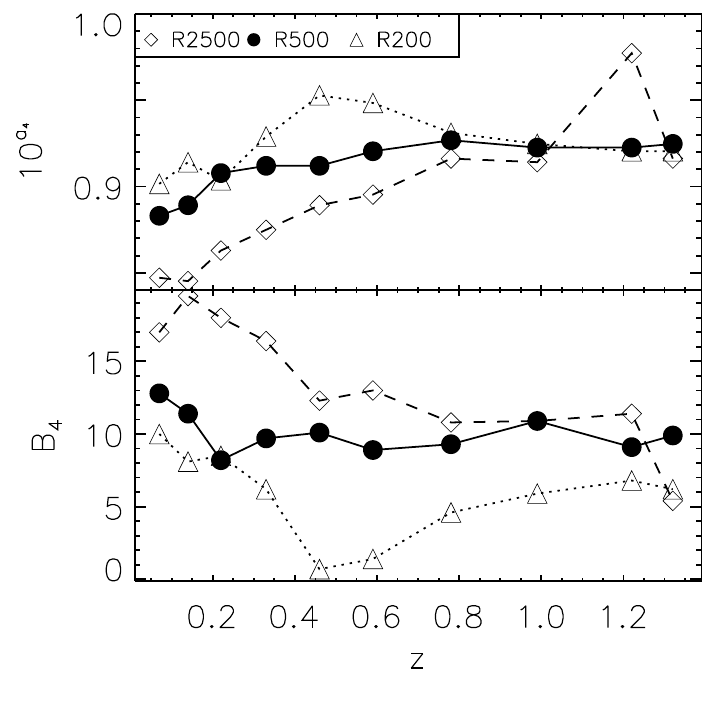}
        \caption{Best-fit parameters of Eq.~\ref{eq:log} (logarithm in the logarithmic plane) for the redshift sub-samples (see Tables~\ref{tab:eq3}).
    The meaning of symbols and lines is similar to Fig.~\ref{fig3}.}
    \label{fig5}
\end{figure}

Fig.~\ref{fig4} shows the parameters of this quadratic form, none of them show a dependence on the redshift at any overdensity. The only exception might appear to be the normalization at $R_{2500}$, however, we notice that $10^{a_3}$ is contained within a narrow range between $0.87$ and $0.91$ without any particular trend for all subsamples with $z<1$. In addition, while at higher redshifts the normalization is significantly higher, in those two epochs the quadratic term is not needed by the data since the corresponding coefficient, $C_3$, fits to values which are approximately $0$ (see Table~\ref{tab:eq2} and red dots in the third panel). With the exclusion of these two instances, the $C_3$ values for $\Delta=2500$ are the most departing from zero. On the other hand, the same values for $\Delta=200$ tend to be consistent with zero within $1\sigma$ for all the subsamples above $z=0.46$ (with the exception of $z=0.78$).

%-----------------------------------
\subsection{Logarithmic form}
\label{sec:log}
%-----------------------------------

The logarithmic form (Eq.~\ref{eq:log}) also accounts for both the drop at low masses and the flattening of the relative fractions to unity at large masses. The advantage with respect to the quadratic expression is that this expression has two parameters similarly to the linear fit. On the other hand, the vertical asymptote is linked to the  
 minimum mass of our cluster sample: $\log(1.27\times 10^{13} h^{-1}/3\times 10^{14})= -1.20$. Regarding this formalism, we also test in Appendix~\ref{app:extra} two more general descriptions in which the constant value of $(-1.20)$ is replaced by the expression $c_4\times |{\rm min}(X)|$  or where the factor $(X+1.2)$ was replaced by $M/M*$ as in \cite{vikhlinin.etal.2009}. 

Some trends with redshift can be seen again at $R_{2500}$ for both parameters. Namely, for the normalization, we find:
\begin{equation}
10^{a_4}=(0.79\pm0.01)+(0.06\pm0.01) (1+zz),
\label{eq:evnormlog}
\end{equation}
while for the curvature we obtain: 
\begin{equation}
b_4=(0.29\pm0.02)-(0.10\pm0.01)(1+z). 
\label{eq:evslopelog}
\end{equation}

\begin{table}
\caption{Means and standard deviations of the 1000 iteration best-fit parameters of Eq.~\ref{eq:glob} for the baryon fraction-mass relation.}             
\label{tab:eq1}      
\centering                         
\begin{tabular}{|c |c c| c c| c |}      
\hline
\multicolumn{3}{|l}{Baryon fraction} & \multicolumn{3}{l|}{$\Delta=2500$} \\
\hline
$z$ & $A_2$ & $\sigma_{A_2}$ & $B_2$ & $\sigma_{B_2}$ &  $10^{a_2}$\\     
\hline                        
0.07 &  -6.2  &    0.2  &    8.9  &    0.8  &    0.87\\
0.14 &  -5.9  &    0.3  &    9.2  &    1.1  &    0.87\\
0.22 &  -5.1  &    0.2  &    7.9  &    0.8  &   0.89\\
0.33 &  -4.5  &    0.3  &    8.4  &    0.9  &    0.90\\
0.46 &  -4.1  &    0.3  &    6.2  &    0.9  &    0.91\\
0.59 &  -3.5  &    0.3  &    7.2  &    0.9  &    0.92\\
0.78 &  -2.0  &    0.8  &    8.1  &    1.3  &    0.95\\
0.99 & -1.6  &    0.5  &    9.0  &    0.9  &     0.96\\
1.22 &   3.1  &   0.8  &  13.6  &   1.0  &   1.07\\
1.32 &  0.2  &    0.5  &   9.3  &    0.7  &    1.00 \\
\hline                                  
\multicolumn{6}{|c|}{$\Delta=500$} \\
\hline
$z$ & $A_2$ & $\sigma_{A_2}$ & $B_2$ & $\sigma_{B_2}$ &  $10^{a_2}$\\     
\hline      
0.07 &   -4.5  &    0.3  &    4.0  &    0.5  &   0.90\\
0.14 &   -4.2  &    0.3  &    3.5  &    0.6  &    0.91\\
0.22 &  -3.5  &    0.2  &    2.4  &    0.5  &    0.92\\
0.33 &   -3.3  &    0.2  &    3.1  &    0.5  &    0.93\\
 0.46 &  -3.3  &    0.1  &    3.6  &    0.3  &    0.93\\
0.59 &  -3.0  &    0.1  &    3.0  &    0.4  &     0.93\\
0.78 &   -2.7  &    0.1  &    3.6  &   0.6  &   0.94\\
0.99 &   -2.7  &    0.2  &    4.5  &   0.6  &   0.94\\
1.22 & -2.9  &    0.2  &    4.2  &    0.6  &    0.94\\
1.32 & -2.4  &    0.3  &    5.3  &    0.7  &    0.95  \\
\hline
\multicolumn{6}{|c|}{$\Delta=200$} \\
\hline
$z$ & $A_2$ & $\sigma_{A_2}$ & $B_2$ & $\sigma_{B_2}$ &  $10^{a_2}$\\    
\hline   
0.07 &  -3.8  &    0.3  &    3.1  &    0.5  &    0.92\\
0.14 &  -3.3  &    0.2  &    2.4  &    0.4  &    0.93\\
0.22 &  -3.6  &    0.3  &    2.2  &    0.5  &    0.92\\
0.33 &  -2.7  &    0.2  &    1.7  &    0.4  &    0.94\\
0.46 &  -2.0  &    0.2  &    0.2  &    0.4  &    0.95\\
0.59 &  -2.2  &    0.1  &    0.4  &    0.4  &    0.95\\
0.78 &  -2.7  &    0.1  &    1.5  &    0.3  &    0.94\\
0.99 &  -3.0  &    0.2  &    2.2  &    0.7  &   0.93\\
1.22 &  -3.0  &    0.1  &    2.9  &    0.4  &    0.93\\
1.32 &  -3.1  &    0.1  &    2.6  &    0.4  &    0.93 \\
\hline                                   
\end{tabular}
\tablefoot{The first column shows the redshift of the subsample, while the last column shows the value of the relative fractions at $M=3\times 10^{14}$\msun. }
\end{table}

\begin{table}
\caption{Means and standard deviations of the 1000 iteration best-fit parameters of Eq.~\ref{eq:stanek} for the baryon fraction-mass relation.}      
\label{tab:eq2}      
\centering      
\begin{tabular}{|c |c c| c c| c c|  c |}   
\hline
\multicolumn{4}{|l}{Baryon fraction} & \multicolumn{4}{l|}{$\Delta=2500$} \\
\hline
$z$ & $A_3$ & $\sigma_{A_{3}}$ & $B_3$ & $\sigma_{B_{3}}$ & $C_3$ & $\sigma_{C_{3}}$ &  $10^{a_{3}}$ \\
\hline   
0.07 & -5.8 & 0.2  &    6.3  &    1.0  &   -5.4  &    1.7  &       0.87\\
0.14 & -5.8 & 0.3  &    8.6  &    1.5  &   {\bf -2.2}  &  {\bf 3.0}  &      0.87\\
0.22 & -4.8 & 0.2  &    6.2  &    0.9  &   -6.0  &    2.0  &        0.90\\
0.33 & -4.4 & 0.2  &    3.7  &    1.3  &   -8.5  &    1.9  &        0.90\\
0.46 & -4.1 & 0.3  &    3.8  &    2.2  &   -4.1  &    3.5  &      0.91\\
0.59 & -4.0 & 0.3  &    1.5  &    1.7  &   -8.2  &    2.3  &        0.91\\
0.78 & -3.7 & 1.2  &   -1.0  &    5.1  &   -8.5  &    4.6  &        0.92\\
0.99 & -4.5 & 0.6  &   -3.8  &    2.1  &   -11.0  &    1.8  &    0.90\\
1.22 &  -1.6   &  2.6 &   0.6   &  6.8  & -8.2  &   4.3 &    1.10 \\
1.32 &  -0.2 & 1.3 &    7.8  &    4.3  &   {\bf -1.1 } &   {\bf  3.0}  &        1.00 \\
\hline        
\multicolumn{8}{|c|}{$\Delta=500$}\\
\hline
$z$ & $A_3$ & $\sigma_{A_3}$ & $B_3$ & $\sigma_{B_3}$ & $C_3$ & $\sigma_{C_3}$ &  $10^{a_3}$\\
\hline   
 0.07 & -4.4  &  0.2 &    5.4  &    0.6  &   -3.1  &    1.0  &        0.90\\
 0.14 & -4.1  &  0.2 &    5.3  &    0.8  &   -3.8 &    1.2  &       0.91 \\
0.22  & -3.6  &  0.2 &    4.5  &    0.8  &   -3.5  &    1.2  &        0.92 \\
0.33  & -3.3  &  0.2 &    3.8  &    0.9  &   {\bf -1.9}  &    {\bf 1.9}  &        0.93 \\
0.46  & -3.3  &  0.1 &    3.7  &    0.3  &   {\bf -0.7}  &   {\bf 0.9}  &        0.93\\
 0.59 & -2.5  &  0.2 &    4.1  &    0.5  &   -5.3  &    1.5  &        0.94\\
 0.78 & -2.6  &  0.1 &    3.3  &    0.6  &   -2.9  &    2.1  &        0.94\\
 0.99 & -2.5  &  0.2 &    2.8  &    0.6  &   -6.1  &    1.7  &        0.94\\
1.22  & -2.8  &  0.2 &    0.3  &    1.0  &   -7.9  &    1.8  &        0.94 \\
 1.32 & -2.8  &  0.3 &    1.2  &    1.8  &   -5.8  &    2.5  &        0.94\\
\hline        
\multicolumn{8}{|c|}{$\Delta=200$}\\
\hline
$z$& $A_3$ & $\sigma_{A_{3}}$ & $B_3$ & $\sigma_{B_{3}}$ & $C_3$ & $\sigma_{C_{3}}$ &  $10^{a_3}$\\ 
\hline   
0.07 & -3.6  &    0.3  &    4.4  &    0.7  &   -2.5  &    1.1  &        0.92\\
0.14 & -3.1  &    0.1  &    4.1  &    0.4  &   -3.3  &    0.6  &        0.93\\
0.22 & -4.2  &    0.3  &    5.9  &    1.5  &   -4.4  &    1.7  &        0.91\\
0.33 & -2.8  &    0.2  &    3.6  &    1.2  &   -2.7  &    1.6  &        0.94\\
0.46 & -2.0  &    0.2  &    0.2  &    0.4  & {\bf 0.0}& {\bf 0.0}   &    0.95\\
0.59 & -2.2  &    0.2  &    0.5  &    0.6  &   {\bf -0.1}  &  {\bf  1.0}  &      0.95 \\
0.78 & -2.6  &    0.1  &    2.0  &    0.4  &   -1.8  &    1.3  &        0.94\\
0.99 & -2.9  &    0.2  &    2.2  &    0.8  &   {\bf -1.3}  &    {\bf 3.1}  &        0.94\\
1.22 & -3.0  &    0.1  &    2.9  &    0.5  & {\bf 0.0}& {\bf 0.0}&        0.93\\
1.32 & -3.1  &    0.1  &    2.2  &    0.6  &   {\bf -1.5}  &   {\bf 1.6}  &        0.93\\
\hline     
\end{tabular}
\tablefoot{The first column shows the redshift of the sample, while the last column shows the value of the relative fractions at $M=3\times 10^{14}$\msun. The bold characters highlight when the parameter $C_3$ is equal or consistent within 1$\sigma$ to zero.}
\end{table}

\begin{table}
\caption{Means and standard deviations of the 1000 iteration best-fit parameters of Eq.~\ref{eq:log}  for the baryon fraction-mass relation.}             
\label{tab:eq3}      
\centering                         
\begin{tabular}{|c |c c| c c| c |}      
\hline
\multicolumn{3}{|l}{Baryon fraction} & \multicolumn{3}{l|}{$\Delta=2500$} \\
\hline
$z$ & $A_4$ & $\sigma_{A_{4}}$ & $B_4$ & $\sigma_{B_{4}}$ &  $N_4$\\    
\hline          
0.07 & -7.2  &     0.2  &    17.0  &     1.2  &    0.87\\
0.14 & -7.3  &     0.3  &    19.5  &     2.5  &    0.88\\
0.22 & -6.4  &     0.2  &    18.0  &     1.4  &    0.89\\
0.33 & -5.8  &     0.2  &    16.4  &     1.2  &    0.90\\
0.46 & -5.1  &     0.2  &    12.3  &     1.6  &    0.91 \\
0.59 & -4.8  &     0.2  &    13.0  &     1.3  &    0.92 \\
0.78 & -3.8  &     0.5  &    10.8  &     1.4  &    0.93 \\
0.99 & -3.9  &     0.2  &    10.9  &     0.6  &    0.93\\
1.22 & -3.4  &   0.5  &   8.4   &  0.8& 1.00 \\
1.32 & -3.8  &     0.5  &     5.4  &     0.9  &   0.93 \\
\hline                                  
\multicolumn{6}{|c|}{$\Delta=500$} \\
\hline
$z$ & $A_4$ & $\sigma_{A_{4}}$ & $B_4$ & $\sigma_{B_{4}}$ &  $N_4$\\     
\hline     
0.07 &    -5.4  &     0.3  &    12.8  &     1.3  &    0.90 \\
0.14 &    -5.1  &     0.3  &    11.4  &     1.6  &   0.91 \\
0.22 &   -4.2  &     0.3  &      8.2  &     1.5  &    0.92\\
0.33 &   -4.0  &     0.3  &      9.7  &     1.6  &    0.93 \\
0.46 &  -4.0  &     0.1  &      10.1  &     0.7  &    0.93 \\
0.59 &   -3.6  &     0.2  &      8.9  &     1.2  &    0.94 \\
0.78 &    -3.3  &     0.1  &     9.3  &     1.4  &    0.94 \\
0.99 &    -3.5  &     0.1  &    10.9  &     1.2  &    0.94 \\
1.22 &  -3.5  &     0.1  &      9.1 &     1.0  &    0.94  \\
1.32 &     -3.4  &     0.2  &     9.9  &     1.2  &    0.94\\
\hline
\multicolumn{6}{|c|}{$\Delta=200$} \\
\hline
$z$ & $A_4$ & $\sigma_{A_{4}}$ & $B_4$ & $\sigma_{B_{4}}$ &  $N_4$\\    
\hline   
0.07 &   -4.5  &     0.3  &     10.0  &     1.4 &    0.92 \\
0.14 &   -3.9  &     0.3  &     8.1  &     1.1  &    0.93 \\
0.22 &   -4.4  &     0.4  &     8.5  &     1.9  &    0.92 \\
0.33 &   -3.2  &     0.3  &     6.2  &     1.4  &    0.94 \\
0.46 &   -2.1  &     0.3  &     0.7  &     1.3  &    0.95\\
0.59 &   -2.3  &     0.2  &     1.4  &     1.2  &    0.95 \\
0.78 &   -3.1  &     0.1  &     4.6  &     0.9  &    0.94 \\
0.99 &   -3.4  &     0.2  &     5.9  &     1.9  &    0.93 \\
1.22 &   -3.6  &     0.1  &     6.8  &     1.2  &    0.93\\
1.32 &   -3.6  &     0.1  &     6.2  &     0.8  &    0.93\\
\hline                                   
\end{tabular}
\tablefoot{The first column shows the redshift of the sample, while the last column shows the value of the relative fractions at $M=3\times 10^{14}$\msun. }
\end{table}

\section{Gas fraction: Tables}
\label{app:gas}

In the Tables~\ref{tab:eq1gas}, \ref{tab:eq2gas}, \ref{tab:eq3gas}, we report the means and standard deviations of the 1000 iteration best-fitting parameters of Eqs.~\ref{eq:glob}, \ref{eq:stanek}, and \ref{eq:log}, respectively, used to fit the relations between the hot-gas fraction and the mass evaluated at $R_{2500}$, $R_{500}$, and $R_{200}$. In the last column, we list  the normalizations of the models at $M=3\times 10^{14}$M$_{\odot}$.

With respect to the results presented in Sect.~\ref{app:bar}. The values of the normalizations are reduced, as expected, for all the three models, but they keep the trends discussed above. In case of the linear fit, the normalizations are almost redshift-independent for the $\Delta=500$ and $200$, and mildly decreasing from high redshift towards $z=0$ for the highest overdensity: 
\begin{equation}
    10^{a_2}=0.58 + 0.12 \times (1+z).
\end{equation}
The slopes of the linear fit of the gas fractions are always steeper than the baryon fraction ones at all overdensities. For the parabola we notice again a large number of instances at $R_{200}$ when the $C_3$ best-fitting values is close or equal to zero, while at higher overdensities they are fewer for the gas fraction with respect to the baryon fraction. In addition, the parameter $b_4$ of the logarithmic fit is always larger than the respective values for the baryon fraction. All these results indicate a more pronounced curvature for gas fraction. 
This result is particularly relevant for the sub-samples that show a more distinct curvature (see the trends shown in Fig.~\ref{fig6}), and it is more incisive in case of the hot-gas fraction with respect to the baryon fraction because of the stronger curvature of the fraction-mass relation (see for example the comparison between the ranges of the $y$-axis in Fig.~\ref{fig7} and Fig.~\ref{fig:zgas}).

In general, as for the baryon fraction, we do not see a significant difference in the performance between the parabolic or logarithmic descriptions, being both able to appropriately describe the data. Although we notice that the logarithmic curvature, in case of the hot-gas, is even more dependent on the redshift: 
\begin{equation}
b_4=(0.44\pm0.02)-(0.13\pm0.01)(1+z).
\end{equation}

\begin{table}
\caption{Means and standard deviations of the 1000 iteration best-fit parameters of Eq.~\ref{eq:glob}  for the hot-gas fraction-mass relation.}             
\label{tab:eq1gas}      
\centering                         
\begin{tabular}{|c |c c| c c| c |}      
\hline
\multicolumn{3}{|l}{Hot-gas fraction} &
\multicolumn{3}{l|}{$\Delta=2500$} \\
\hline
$z$ & $A_2$ & $\sigma_{A_2}$ & $B_2$ & $\sigma_{B_2}$ &  $10^{a_2}$\\     
\hline                        
0.07 & -14.7   &  0.4  &  17.4   &  1.5 &  0.71\\
0.14 & -14.0   &  0.4  &  16.3   &  1.5 &  0.72\\
0.22 & -13.1   &  0.2  &  11.7   &  0.9 &  0.74\\
0.33 & -12.6   &  0.4  &  14.2   &  1.4 &  0.75\\
0.46 & -12.0   &  0.5  &  12.6   &  1.4 &  0.76\\
0.59 & -10.8   &  0.4  &  13.4   &  0.8 & 0.78\\
0.78 &  -9.8   &  0.9  &  14.6   &  1.5 &  0.80\\
0.99 & -10.1   &  0.9  &  14.6   &  1.7 &  0.79\\
1.22 &  -3.4   &  1.3  &  23.4   &  1.8 &  0.92\\
1.32 &  -4.3   &  1.4  &  21.1   &  1.7 &  0.91\\
\hline                                  
\multicolumn{6}{|c|}{$\Delta=500$} \\
\hline
$z$ & $A_2$ & $\sigma_{A_2}$ & $B_2$ & $\sigma_{B_2}$ &  $10^{a_2}$\\     
\hline      
0.07 &  -9.8   & 0.3   &  5.1  &  0.6 & 0.80\\
0.14 &  -9.5   & 0.3   &  4.6  &  0.8 & 0.80\\
0.22 &  -8.7   & 0.2   &  3.6  &  0.6 &  0.82\\
0.33 &  -8.6   & 0.2   &  4.4  &  0.5 &  0.82\\
0.46 &  -8.7   & 0.1   &  4.8  &  0.3 &  0.82\\
0.59 &  -8.6   & 0.2   &  4.5  &  0.5 & 0.82\\
0.78 &  -8.7   & 0.2   &  5.8  &  0.7 &  0.82\\
0.99 &  -9.0   & 0.2   &  5.9  &  0.7 &  0.81\\
1.22 &  -8.8   & 0.2   &  7.7  &  0.6 &  0.82\\
1.32 &  -8.6   & 0.4   &  8.3  &  0.9 &  0.82\\
\hline
\multicolumn{6}{|c|}{$\Delta=200$} \\
\hline
$z$ & $A_2$ & $\sigma_{A_2}$ & $B_2$ & $\sigma_{B_2}$ &  $10^{a_2}$\\    
\hline   
0.07 &   -8.5  & 0.3  &  4.2 &  0.6  &  0.82 \\
0.14 &   -8.0  & 0.2  &  3.7 &  0.5  &  0.83 \\
0.22 &   -6.9 &  0.2  &  1.5 &  0.4  & 0.85 \\
0.33 &   -7.1  & 0.2  &  2.2 &  0.4  &  0.85 \\
0.46 &   -6.5  & 0.2  &  0.6 &  0.4  &  0.86 \\
0.59 &   -7.2  & 0.1  &  1.5 &  0.3  &  0.85 \\
0.78 &   -7.9  & 0.1  &  3.2 &  0.3  &  0.83 \\
0.99 &   -8.5  & 0.2  &  3.2 &  0.6  &  0.82 \\
1.22 &   -8.5  & 0.1  &  6.3 &  0.5  &  0.82 \\
1.32 &   -8.7  & 0.2  &  6.1 &  0.7  & 0.82 \\  
\hline                                   
\end{tabular}
\tablefoot{The first column shows the redshift of the subsample, while the last column shows the value of the relative fractions at $M=3\times 10^{14}$\msun. }
\end{table}

\begin{table}
\caption{Means and standard deviations of the 1000 iteration best-fit parameters of Eq.~\ref{eq:stanek}  for the hot-gas fraction-mass relation.}      
\label{tab:eq2gas}      
\centering      
\begin{tabular}{|c |c c| c c| c c| c| c |}   
\hline
\multicolumn{4}{|l}{Hot-gas fraction} & \multicolumn{4}{l|}{$\Delta=2500$} \\
\hline
$z$ & $A_3$ & $\sigma_{A_3}$ & $B_3$ & $\sigma_{B_3}$ & $C_3$ & $\sigma_{C_3}$ &  $10^{a_3}$ \\
\hline
0.07 &  -14.0 &  0.3 &   10.6  &    2.0  &  -12.2  &    3.0  &    0.72 \\
0.14 &  -13.5 &  0.4 &   13.5  &    1.9  &   -7.2  &    3.5  &    0.73 \\
0.22 &  -12.7 &  0.3 &   10.2  &    1.1  &   -5.7  &    2.7  &      0.75 \\
0.33 &  -12.2 &  0.3 &    7.5  &    1.7  &  -12.7  &    2.6  &      0.76\\
0.46 &  -12.2 &  0.5  &   6.8  &    3.2  &  -10.0  &    5.1  &     0.76\\
0.59 &  -11.3 &  0.3 &    7.1  &    1.9  &  -7.6  &    2.1  &     0.77 \\
0.78 &  -12.5 &  1.1  &  -0.8  &    4.9  &  -14.9  &    4.6  &     0.75 \\
0.99 &  -13.8 &  0.8 &   -7.0  &    3.5  &  -20.9  &    3.2  &    0.73 \\
1.22 &   -5.3 &  3.5  &  17.0  &    11.2  &  -4.6  &    8.0  &     0.87 \\
1.32 &  -10.1 &  3.1  &  3.4  &    8.7  &  -12.0  &    5.8  &    0.79 \\

\hline                                   
\multicolumn{8}{|c|}{$\Delta=500$} \\
\hline
$z$ & $A_3$ & $\sigma_{A_3}$ & $B_3$ & $\sigma_{B_3}$ & $C_3$ & $\sigma_{C_3}$ &  $10^{a_3}$ \\
\hline
0.07 & -9.7  &    0.3  &  6.9  &    0.9  &   -3.4  &    1.3  &   0.80\\
0.14 & -9.5  &    0.2  &  7.3  &    0.7  &   -5.6  &    1.1  &    0.80\\
0.22 & -8.9  &    0.3  &  5.9  &    1.4  &   -3.3  &    1.9  &     0.81 \\
0.33 & -8.7  &    0.2  &  5.9  &    0.9  &   -3.2  &    1.7  &    0.82 \\
0.46 & -8.7  &    0.1  &  4.9  &    0.4  & {\bf 0.0}  & {\bf 0.0} & 0.82 \\
0.59 & -8.3  &    0.2  &  5.7  &    0.6  &   -4.5  &    1.5    & 0.83 \\
0.78 & -8.5  &    0.2  &  5.2  &    0.8  &   -3.2  &    2.4    & 0.82 \\
0.99 & -8.6  &    0.3  &  5.1  &    0.7  &   -5.1  &    2.5  &    0.82 \\
1.22 & -8.7  &    0.2  &  5.4  &    1.1  &   -5.3  &    2.3  &     0.82 \\
1.32 & -8.7  &    0.3  &  4.5  &    2.2  &   -6.9  &    3.6  &     0.82 \\
\hline                                   
\multicolumn{8}{|c|}{$\Delta=200$} \\
\hline
$z$ & $A_3$ & $\sigma_{A_3}$ & $B_3$ & $\sigma_{B_3}$ & $C_3$ & $\sigma_{C_3}$ &  $10^{a_3}$ \\
\hline
0.07 & -8.3  &   0.3  &    5.9  &    0.9  &   -3.4  &    1.4  &     0.83\\
0.14 & -7.7  &   0.1  &    5.8  &    0.4  &   -4.4  &    0.7  &    0.84\\
0.22 & -7.1  &   0.2  &    2.5  &    1.1  &   {\bf -1.3 } &   {\bf 1.3}  &     0.85 \\
0.33 & -7.4  &   0.2  &    4.6  &    1.1  &   -3.4  &    1.5  &     0.84 \\
0.46 & -6.5  &   0.2  &    0.6  &    0.4  &{\bf 0.0}& {\bf 0.0}&      0.86 \\
0.59 & -7.2  &   0.1  &    1.5  &    0.3  &{\bf 0.0}& {\bf 0.0}&   0.85 \\
0.78 & -7.7  &   0.1  &    3.6  &    0.3  &   -2.6  &   1.0  &      0.84 \\
0.99 & -8.3  &   0.2  &    3.5  &    0.7  &   {\bf -2.8}  &   {\bf 2.8}  &      0.83 \\
1.22 & -8.5  &   0.1  &    6.3  &    0.6  & {\bf 0.0}&{\bf 0.0}&  0.82 \\
1.32 & -8.7  &   0.2  &    5.7  &    1.0  &   {\bf -2.2}  &  {\bf  3.2}  &      0.82 \\
\hline                                   
\end{tabular}
\tablefoot{The first column shows the redshift of the sample, while the last column shows the value of the relative fractions at $M=3\times 10^{14}$\msun. The bold characters highlight when the parameter $C_3$ is equal or consistent within 1$\sigma$ to zero.}
\end{table}

\begin{table}
\caption{Means and standard deviations of the 1000 iteration best-fit parameters of Eq.~\ref{eq:log} for the hot-gas fraction-mass relation.}             
\label{tab:eq3gas}      
\centering                         
\begin{tabular}{|c |c c| c c|  c |}      
\hline
\multicolumn{3}{|l}{Hot-gas fraction} & \multicolumn{3}{l|}{$\Delta=2500$} \\
\hline
$z$ & $A_4$ & $\sigma_{A_{4}}$ & $B_4$ & $\sigma_{B_{4}}$ &  $N_4$\\    
\hline          
0.07 &   -16.8  &     0.3  &    32.3  &     2.0  &  0.72 \\
0.14 &   -16.1  &     0.4  &    31.9  &     2.7  &  0.73 \\
0.22 &   -14.9  &     0.2  &    26.5  &     1.9  &   0.74 \\
0.33 &   -14.6  &     0.2  &    27.2  &     1.7  &   0.75 \\
0.46 &   -14.2  &     0.3  &    25.2  &     2.4  &   0.76 \\
0.59 &   -13.2  &     0.3  &    20.9  &     1.0  &   0.77 \\
0.78 &   -13.0  &     0.4  &    20.2  &     1.3  &   0.77 \\
0.99 &   -13.3  &     0.3  &    19.7  &     1.1  &   0.76 \\
1.22 &   -12.5  &     1.0  &    18.1  &     2.1  &   0.78 \\
1.32 &   -13.7  &     0.9  &    14.3  &     1.5  &   0.75\\

\hline                         
\multicolumn{6}{|c|}{$\Delta=500$} \\
\hline
$z$ & $A_4$ & $\sigma_{A_{4}}$ & $B_4$ & $\sigma_{B_{4}}$ &  $N_4$\\     
\hline     
 0.07 &   -11.1  &     0.4  &    16.8  &     1.7&  0.80 \\
 0.14 &   -10.7  &     0.4  &    15.3  &     1.9&   0.80 \\
 0.22 &    -9.8  &     0.4  &    12.7  &     1.8&   0.82 \\
 0.33 &    -9.8  &     0.3  &    14.3  &     1.5&   0.82 \\
 0.46 &    -9.7  &     0.2  &    14.4  &     1.2&   0.82 \\
 0.59 &    -9.5  &    0.2  &    13.5  &     1.3 &   0.82 \\
 0.78 &    -9.7  &    0.2  &    14.3  &     1.6 &   0.82 \\
 0.99 &   -10.0  &    0.2  &    14.9  &     1.5 &   0.82 \\
 1.22 &   -10.1   &   0.1   &     16.7 &     1.0 &  0.82\\
 1.32 &   -10.0  &    0.2  &    16.8  &     1.7 &   0.82 \\
\hline
\multicolumn{6}{|c|}{$\Delta=200$} \\
\hline
$z$ & $A_4$ & $\sigma_{A_{4}}$ & $B_4$ & $\sigma_{B_{4}}$ &  $N_4$\\    
\hline   
0.07 &    -9.5  &    0.4  &     13.7  &    1.6  &   0.82 \\
0.14 &    -8.8  &     0.3  &    12.2  &    1.1  &   0.83 \\
0.22 &    -7.4  &     0.3  &     5.3  &    1.4  &   0.85 \\
0.33 &    -7.8  &     0.3  &     8.0  &    1.5  &   0.85 \\
0.46 &    -6.6  &     0.3  &     1.9  &    1.4  &   0.86 \\
0.59 &    -7.5  &     0.2  &     5.0  &    0.9  &   0.85 \\
0.78 &    -8.6  &     0.1  &     9.2  &    0.8  &   0.83 \\
0.99 &    -9.1  &     0.2  &     9.2  &     1.7  &  0.82 \\
1.22 &    -9.6  &     0.2  &    15.4  &     1.5  &  0.82 \\
1.32 &    -9.8  &     0.2  &    15.0  &     1.6  &  0.82 \\
\hline                                   
\end{tabular}
\tablefoot{The first column shows the redshift of the subsample, while the last column shows the value of the relative fractions at $M=3\times 10^{14}$\msun. }
\end{table}

\section{Evolution parametrization}
\label{app:evpar}
\begin{figure}
    \centering
        \includegraphics[width=0.4\textwidth]{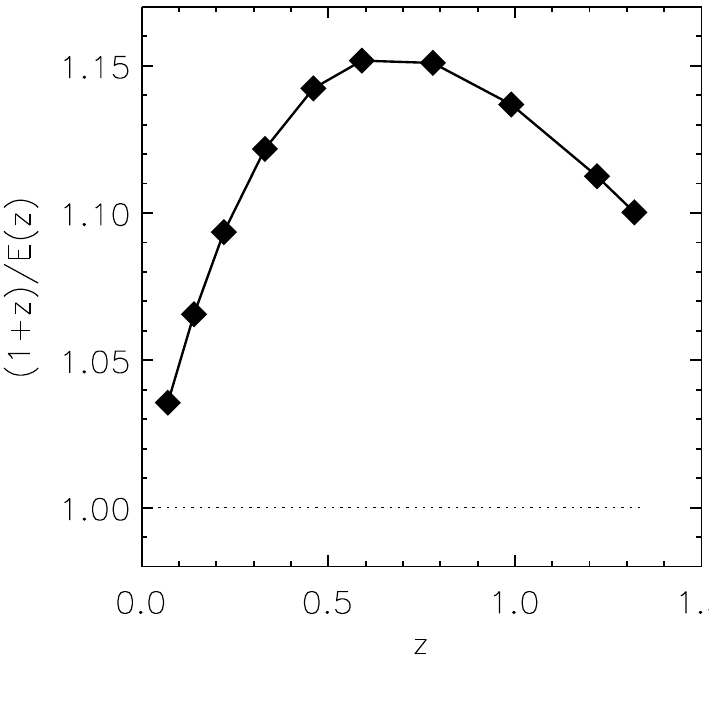}
    \caption{ Ratio between the two parametrizations of the evolution, $(1+z)$ and $E(z)$ in red, as a function of the redshifts of the subsamples used in this work.}
    \label{fig:evolratio}
\end{figure}

In Fig.~\ref{fig:evolratio} we compare the two parametrizations of the evolution that we studied in this work: $(1+z)$ and $E(z)$. The maximum difference between the two curves is around 15 percent which is within the scatter of both baryon and hot gas fractions for the overdensity $\Delta=2500$ (see Fig.~\ref{fig:barpresentation}), where a clear sign of evolution is detected. This comparison explains the reason why the results from the two evolution parametrizations are consistent within each other when a shift along the normalization is considered.

\section{Extra fitting procedures}
\label{app:c}

\subsection{$D$ parameters using $E(z)$}
\label{app:dforez}

In Table~\ref{tab:dforez} we report the values of the $D$ parameters when we use the $E(z)$ parametrization and Eq.~\ref{eq:extx} for both the baryon and gas fractions.

\begin{table}
\caption{Means and standard deviations of the 1000 iteration best-fit parameter $D$ for Eq.~\ref{eq:glob}, Eq.~\ref{eq:stanek}, and Eq.~\ref{eq:log} with the $E(z)$ redshift dependence as in Eq.~\ref{eq:extx}. }    
\label{tab:dforez}      
\centering      
\begin{tabular}{|c |c c| c c| c c|}   
\hline
Baryon  & \multicolumn{2}{c|}{$\Delta=2500$}  & \multicolumn{2}{c|}{$\Delta=500$} & \multicolumn{2}{c|}{$\Delta=200$} \\    
\hline 
Eq.~\ref{eq:glob}:$D_{2,z} \ \sigma_D$ & 224.5 & 5.9 & 123.3  & 5.9 & -34.9 & 10.4  \\ 
Eq.~\ref{eq:stanek}: $D_{3,z}\ \sigma_D$ & 220.7 & 3.5 & 94.3   &   20.8   & -45.3  &  14.8   \\
Eq.~\ref{eq:log}: $D_{4,z} \ \sigma_D$ & 217.8 & 6.8 & 100.7  & 8.4 & -78.0 & 27.4  \\
\hline
\hline
Hot-gas  & \multicolumn{2}{c|}{$\Delta=2500$}  & \multicolumn{2}{c|}{$\Delta=500$} & \multicolumn{2}{c|}{$\Delta=200$} \\    
\hline 
Eq.~\ref{eq:glob}:$D_{2,z} \ \sigma_D$ & 112.8 &3.1  & 12.1  &24.9  &-148.7  & 3.7 \\ 
Eq.~\ref{eq:stanek}: $D_{3,z}\ \sigma_D$ & 88.4& 8.0 &  13.3  & 8.3    &-143.5   & 10.4    \\
Eq.~\ref{eq:log}: $D_{4,z} \ \sigma_D$ & 61.2 & 37.8 & -7.2  & 13.5 & -187.2 & 27.7 \\
\hline
%%%%%
\end{tabular}
\end{table}

\subsection{Procedure on independent subsamples}
\label{app:verify}
We verify whether the hot-gas results on the combined sample could be affected by possible interdependence of the cluster samples extracted from nearby redshifts by
restricting the analysis to $z=0.07$, $z=0.46$ and $z=1.32$. In other words, we combine together three sets of data that can be considered independent because the three redshifts are separated by more than 4 Gyr, which is the typical relaxation time needed for massive structures to relax after a major merger \citep{gianfagna.etal.2023}. In addition, comparing the lowest and highest redshift samples, 98 percent of the low-redshift objects have a formation redshift \citep{darragh-ford.etal.2023} smaller than 1.3 and 35 percent do not have a progenitor in the high-redshift sample. The parameters of the parabola and the logarithmic forms obtained with the reduced sample are in agreement with those of Table~\ref{tab:zgas} within $1\sigma$ for the three overdensities, demonstrating the robustness of our results. 

\subsection{Alternative expressions}
\label{app:extra}
\begin{figure}
    \centering
    \includegraphics[width=0.49\textwidth]{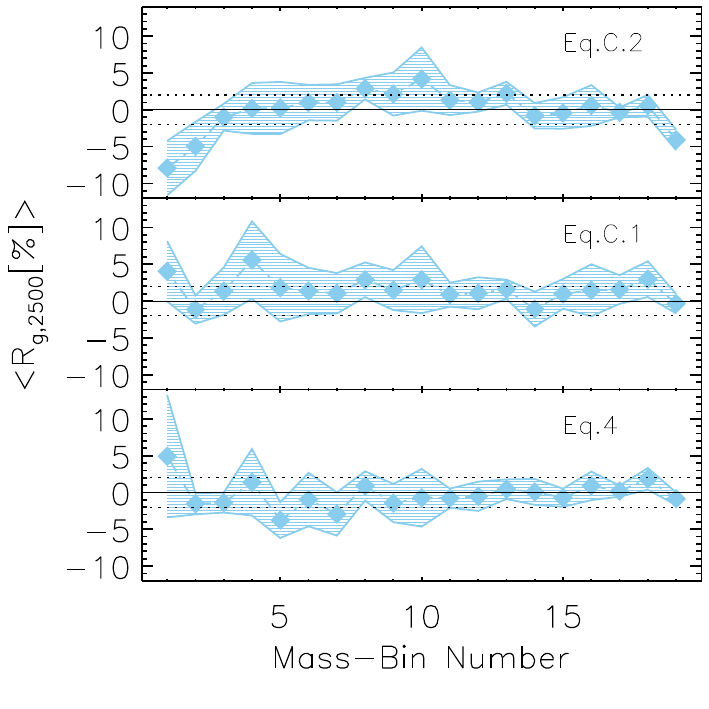}
    \caption{ Average and 1$\sigma$ deviation of the residuals of the hot-gas fraction at $R_{2500}$ at fixed mass-bin number for Eq.~\ref{eq:log}, Eq.~\ref{eq:c1}, and Eq.~\ref{eq:c2}}
    \label{fig:resextra}
\end{figure}

We test a few variations of the logarithmic functional form (Eq.\ref{eq:log}) and compare the averaged residual for the hot-gas fraction evaluated within $R_{2500}$ in Fig.~\ref{fig:resextra}:
\begin{enumerate}
\item A generalization of Eq.~\ref{eq:log} where we added an extra parameter is tested on the sub-samples at fixed redshift:
\begin{equation}
Y=a_4+b_4\log(X+c_4 \times |\min(X)|).
\label{eq:c1}
\end{equation}
The residuals (middle panel) tend to have an higher dispersion than those of Eq.\ref{eq:log} (bottom panel) but do not show extreme deviation for the lowest and highest mass bins. Nevertheless, since this equation foresees an extra parameter it is less desirable. 
\item In \cite{vikhlinin.etal.2009} to fit the hot-gas fraction versus the total mass at $R_{500}$, the latter was re-normalized by $M_*$, the mass scale corresponding to a r.m.s. value of the linear fluctuation amplitude of 1.686, i.e. the linearly extrapolated virial density contrast predicted by the spherical collapse model. This characteristic mass in out sample varies from $M_*=3.7 \times 10^{12}M_\odot$ at $z=0.07$, to $M_*=6.1 \times 10^{10}M_\odot$ at $z=1.32$. We test the following expression:
\begin{equation}
Y=a_4+b_4\times\log(M_{500}/M_*),
\label{eq:c2}
\end{equation}
against the subsamples at fixed redshift. This expression provides residuals (see top panel of Fig.~\ref{fig:resextra}) with deviation in the first bins and in the last, thus, is less desirable than Eq.~\ref{eq:log}.
\end{enumerate}

\section{Baryon and gas fraction medians}
\label{app:med}

\begin{figure}
    \centering
    \includegraphics[width=0.49\textwidth]{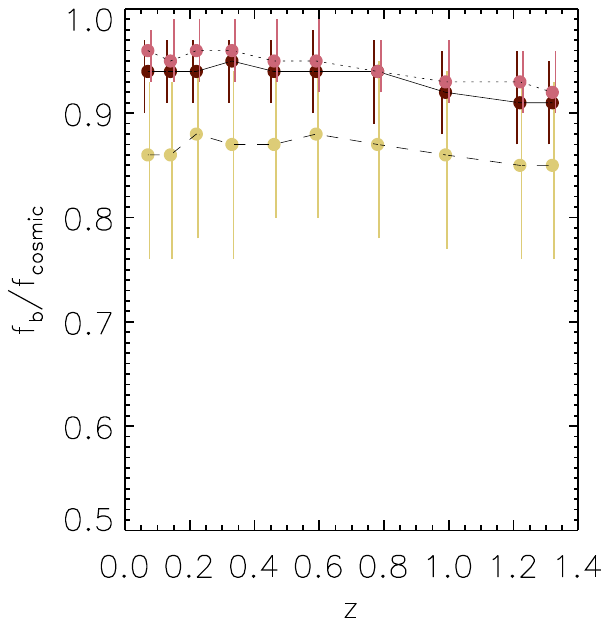}
    \includegraphics[width=0.49\textwidth]{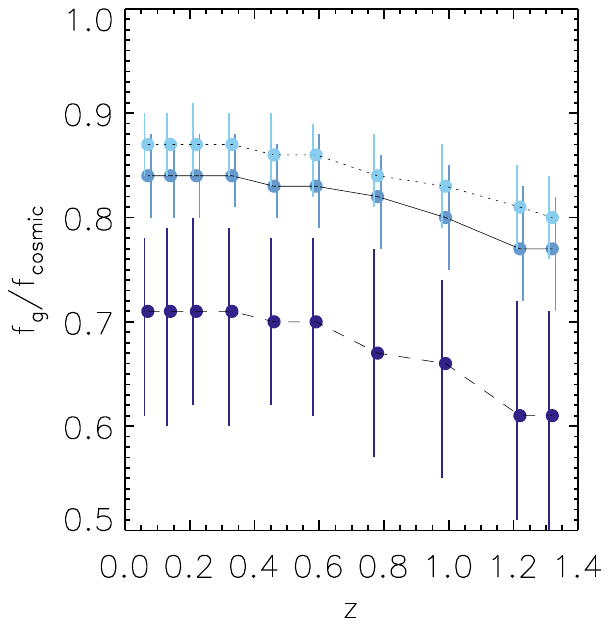}
    \caption{ Median values of the relative baryon fractions (top panel) and gas fractions (bottom panel) as function of redshift computed at $R_{2500}$ (in mustard and navy colors)), $R_{500}$ (in dark-red and medium-blue colors), and $R_{200}$ (in salmon and light-blue colors). The error bars represent the $16^{\rm th}$ and $84^{\rm th}$ percentiles.}
    \label{fig:med}
\end{figure}

\subsection{Baryon fraction medians}
\label{app:medbar}
While the goal of the paper is to parametrize the relation between the relative fraction and the total mass of the systems, it is worth also commenting on the evolution of the baryon-fraction medians of all clusters in each redshift subsamples. The relative figure, top panel of Fig.~\ref{fig:med}, shows that the median values of the baryon fractions do not vary over time at any overdensity.

\subsection{Gas Fraction sample medians}
\label{app:gasmedian}
The bottom panel of Fig.~\ref{fig:med} shows the evolutionary trends for the medians of the hot-gas fractions of all clusters in each redshift subsample.
We can notice an increase of 15, 10, and 9 percent from $z=1.3$ to $z=0.07$ for the overdensities of $R_{2500}$, $R_{500}$ $R_{200}$, respectively. While the values of the extreme redshifts are consistent within the dispersion of the respective subsamples, it is worth noticing that the increase of the median fraction is systematic throughout time.  

Another aspect to remark is that at $z=0.07$, the median values of the relative hot gas fraction at the three overdensities $\Delta = 2500, 500$ and 200, are $0.71$, $0.84$, $0.87$, respectively. The same values for the total baryon content are $0.86$, $0.94$, $0.96$, at the same three overdensities. Therefore, the radius growth for the two fractions has quite a different rate, being the difference between $R_{2500}$ and $R_{200}$ of order of 11 percent for the baryon fraction and 22 percent for the hot-gas fraction. This again corroborates that the depletion of the hot-gas fraction is larger in the innermost radius. 

\subsubsection{Massive systems}
\label{app:massive}
The evolution of the median values of the hot gas fraction disappears if we restrict the analysis to only massive clusters at all redshifts, such as $M_{500}>2\times10^{14}$M$_{\odot}$, which is a condition often seen in observational samples (see Sect.~\ref{sec:massive}).
In this case, at $z=0$ the hot-gas fraction at $R_{500}$ is equal to 0.13, corresponding to a relative fraction of 0.84 \citep[see also discussion in][]{eckert.etal.2019} and the same value is also found for the samples at $z\sim 0.5$ and $z\sim 1$. No evolution is also detected for the values of the 16$^{\rm th}$ and 84$^{\rm th}$ percentiles, which are equal to 0.124 and 0.136, respectively. 

The same massive clusters return a median value of the gas fraction computed within the radial interval from 0.8 to 1.2 $R_{2500}$ of 0.136 (being 0.123 and 0.149 the $16^{\rm th}$ and $84^{\rm th}$ percentiles respectively), in very good agreement with the measurements by \citet{Mantz.etal.2022}, equal to 0.133 (with 0.112 and 0.152 as percentiles).

\end{appendix}
\end{document}